\documentclass[prb,aps,twocolumn,showpacs]{revtex4}
\usepackage{graphics}
\usepackage{graphicx}
\usepackage{amssymb}
\usepackage{amsmath}
\usepackage{bm}

\newcommand{\mytitle}[1]{
 \twocolumn[\hsize\textwidth\columnwidth\hsize
 \csname@twocolumnfalse\endcsname #1 \vspace{1mm}]}

\newcommand{\beq}{\begin{equation}}
\newcommand{\eeq}{\end{equation}}
\newcommand{\bea}{\begin{eqnarray}}
\newcommand{\eea}{\end{eqnarray}}

\begin{document}

\title{Non-equilibrium quantum transport through a dissipative resonant level}

\author{Chung-Hou Chung$^{1,2}$, Karyn Le Hur$^{3,4}$, Gleb Finkelstein$^{5}$, Matthias Vojta$^{6}$, and Peter W\" olfle$^{7,8}$}
\affiliation{$^{1}$Department of Electrophysics, National Chiao-Tung University, HsinChu, Taiwan, R.O.C. \\
$^{2}$ Physics Division, National Center for Theoretical Sciences, HsinChu, Taiwan, R.O.C. \\
$^{3}$ Center for Theoretical Physics
Ecole Polytechnique and CNRS
91128 Palaiseau, France\\
$^{4}$ Department of Physics and Applied Physics, Yale University, New 
Haven, CT, USA \\
$^{5}$ Department of Physics, Duke University, Durham, NC 27708, U.S.A.\\
$^{6}$ Institut f\"ur Theoretische Physik, 
Technische Universit\"at Dresden, 01062 Dresden, Germany\\
$^{7}$Institut f\"ur Theorie der Kondensierten Materie, 
KIT, 76128 Karlsruhe, Germany\\
$^{8}$ Institut f\"ur Nanotechnologie, KIT, 76021 Karlsruhe, Germany} 
\date{\today}

\begin{abstract}
The resonant-level model represents a paradigmatic quantum  
system which serves as a basis for many other quantum impurity models. 
We provide a comprehensive analysis of 
the non-equilibrium transport near a quantum phase transition 
in a spinless dissipative resonant-level model, extending earlier work 
[Phys. Rev. Lett. {\bf 102}, 216803 (2009)]. 
A detailed derivation of a rigorous mapping of our system 
onto an effective Kondo model is presented. 
A controlled energy-dependent 
renormalization group approach is applied to compute the non-equilibrium 
current in the presence of a finite bias voltage $V$. In the linear response 
regime $V\rightarrow 0$, 
the system exhibits as a function of the dissipative strength 
a localized-delocalized quantum transition of the 
Kosterlitz-Thouless (KT) type. We address fundamental issues of the 
non-equilibrium transport near the quantum phase transition: Does the 
bias voltage play the same role 
as temperature to smear out the transition? What is the scaling of the 
non-equilibrium conductance near the transition? 
At finite temperatures, we show that the conductance 
follows the equilibrium scaling for $V< T$, while it obeys a 
distinct non-equilibrium profile for $V>T$. We furthermore provide new 
signatures of the transition in the finite-frequency current noise and AC 
conductance via the recently developed Functional 
Renormalization Group (FRG) approach. 
The generalization of our analysis to non-equilibrium 
transport through a resonant level coupled to two 
chiral Luttinger-liquid leads, generated by the fractional quantum Hall 
edge states, is discussed. Our work 
on dissipative resonant level has direct relevance to 
the experiments in a quantum dot coupled to resistive environment, 
such as H. Mebrahtu {\it et al.}, Nature {\bf 488}, 61, (2012).

\end{abstract}

\maketitle

\section{\bf Introduction}

Quantum phase transitions (QPTs)\cite{sachdevQPT,Steve} 
 which separate competing ground states represent 
generic phenomena in solid-state systems at zero temperature. 
The transition is frequently found to be continuous, 
often times giving rise to a quantum critical point. In the neighborhood of a 
quantum critical point of a metallic system the finite temperature 
properties as a rule show  
non-Fermi liquid behavior\cite{vojtaRMP}. 
In recent years, quantum phase transitions at the nanoscale have 
attracted much attention\cite{saleur,Ingold,nazarov-book,lehur1,lehur2,zarand,Markus,matveev,gefen,buttiker,Zarand2,zoller}. Much of the effort has been focused on the 
breakdown of the Kondo effect in transport of a quantum dot due to its 
coupling to a dissipative environment. 
However, relatively less is known about the corresponding 
out-of-equilibrium 
properties\cite{chung1,chung2,latha,Feldman,mitra,Goldhaber,QMSi,ybkim,zwergerbook,florens}. 
A finite bias voltage applied across a nanosystem is 
expected to smear out the equilibrium transition, 
but the current-induced decoherence 
might act quite differently as compared to 
thermal decoherence at finite temperature $T$, 
resulting in exotic behavior near the transition. 

Meanwhile, understanding the interplay of electron interactions and 
non-equilibrium effects in quantum systems is one of the 
most challenging open questions in condensed matter physics. 
Many of the theoretical approaches that have been proven 
so successful in treating strongly correlated systems in equilibrium 
are simply inadequate once the system is out of equilibrium. 
The real-time Schwinger-Keldysh formalism\cite{keldysh} has been known 
as the most successful approach to non-equilibrium dynamics since 
it offers a controlled perturbative expansion of the density operator. 
However, care must be taken to avoid the appearance of infrared divergences,
in the perturbative approaches. Though much is known 
for quantum impurity systems in equilibrium, understanding 
 their properties in non-equilibrium steady-state is still 
limited. Nevertheless, significant progress has been made   
by different approaches, such as (1) analytical approximations: perturbative 
renormalization group method (RG) \cite{noneqRG,schoellerRG}, 
Hamiltonian flow equations\cite{kehrein}, Functional RG\cite{meden,FRG}, 
strong-coupling expansions\cite{mora}, master equations\cite{timm}; 
(2) exact analytical solutions: field theory techniques\cite{fendley}, 
the scattering Bethe Ansatz\cite{andrie}, mapping of a steady-state non-equilibrium problem onto an effective equilibrium system\cite{hur-eq-noneq}; (3) numerical methods: 
time-dependent density matrix renormalization group (RG)\cite{saleur-dagotto}, 
time-dependent numerical RG\cite{anders}, 
diagrammatic Monte Carlo\cite{fabrizio}, and   
imaginary-time nonequilibrium quantum Monte Carlo\cite{han}.

In this paper,  
we provide a comprehensive analysis of 
the non-equilibrium transport near a quantum phase transition 
in a dissipative resonant level model by employing the 
recently developed frequency-dependent RG\cite{noneqRG} 
and Functional RG approaches\cite{FRG}, and extending our earlier work in 
Ref.~\onlinecite{chung1}. 
We aim to address several fundamental questions related 
to the non-equilibrium transport in quantum dot settings, 
such as: what is the distinct non-equilibrium conductance profile 
at zero temperature compared to that in equilibrium at finite temperatures 
near the transition? is there any scaling  behavior of the conductance at 
finite temperatures and finite bias voltage near the transition?

For this purpose, we investigate three classes of 
typical nano-models comprising a spinless resonant level coupled to:
(i) two non-interacting Fermi-liquid leads subject to 
an Ohmic dissipative environment, where an Ohmic environment 
can be realized in a nanoscale resistor 
and has many applications in physics ranging from mesoscopic 
physics (Refs. \onlinecite{lehur2,Markus,matveev}) 
to biological systems\cite{Mckenzie2}, 
(ii) two interacting fermion baths, in particular 
two Fractional Quantum Hall Edge (FQHE)\cite{FQHE}  
leads, or the ``chiral Luttinger liquids'' where electrons 
on the edge of a 2D fractional quantum Hall system 
show one-dimensional chiral Luttinger liquid behaviors 
with only one species of electrons (left or right movers), 
(iii) two interacting Luttinger-liquid leads subject to 
an Ohmic dissipative environment.

In the class (i) model, 
the QPT separating the conducting and insulating phase 
for the level is solely driven by dissipation, which can be modeled by a 
bosonic bath. Dissipation-driven QPTs have been 
addressed theoretically and experimentally in various systems, such as: 
quantum dot systems\cite{zarand,finkelstein}, 
Josephson junction arrays\cite{Josephson,demler,lobos}, 
superconducting thin film\cite{Kapitulnik,zimanyi}, superconducting 
qubit\cite{Rimberg}, qubits or resonant level systems coupled 
to photonic cavities\cite{Kontos,cavities}, and biological 
systems\cite{Mckenzie2,McKenzie}. Here, we focus on 
the non-equilibrium properties of the system near 
quantum phase transition. Meanwhile, for the class (ii) model, tunneling 
of electrons or quasi-particles between two FQHE states may in general 
suffer from the electron-electron interactions in FQHE. Interesting 
experimentally relevant questions arise regarding how interaction 
effects modify the 
nonequilibrium charge transport in such systems.   
Furthermore, one can extend the above two classes of models to 
a more general class (iii) model where both electron-electron interactions 
and the dissipation are present in the FQHE setpups, which have not been 
explored both theoretically and experimentally. Our results have relevance 
for the recent experiment in Ref.~\onlinecite{finkelstein} where the electronic 
transport through a resonant level in a nanotube exhibits the 
Luttinger liquid behavior, namely the conductance demonstrates a non-trivial 
power-law suppression as a function of bias voltage.

This paper is organized as follows: In Section II A, the model Hamiltonian 
of class (i) is introduced. In Section II B, we establish rigorious 
mappings of our model system, the class (i) model at a finite bias voltage, 
onto the out-of-equilibrium anisotropic Kondo model as well as onto 
class (ii) and (iii) model systems subject to a finite voltage bias. 
We compute the current operator 
in Section II C for these three classes of models.
We employ the nonequilibrium 
RG approach in Section III. Our results on nonequilibrium transport 
near the quantum phase transition both at zero and finite temperatures 
are presented in Section IV, followed by the results on the nonequilibrium 
finite-frequency current noise in Section V. We make a few remarks on the 
important issues of nonequilibrium quantum criticality 
in Section VI. Finally, we draw conclusions in Section VII.

\section{Model Hamiltonian}

\subsection{Dissipative resonant level model}

Our Hamiltonian in all of the three classes of models mentioned above  
takes the following generic form:
\begin{eqnarray}
H &=& \sum_{k,i=1,2} [\epsilon(k)-\mu_i] c^{\dagger}_{k i}c_{k i} + t_{i}
c^{\dagger}_{ki} d + h.c. \\
&+& \sum_{r} \lambda_{r} (d^{\dagger}d-1/2) (b_{r} + b^{\dagger}_{r}) +
\sum_{r} \omega_{r} b^{\dagger}_{r} b_{r},  \nonumber
\label{RLM}
\end{eqnarray}
where $t_{i}$ is the (real-valued) hopping amplitude between the 
lead $i$ and the quantum dot, $c_{ki}$ and $d$ are electron operators for 
the (Fermi-liquid type) leads and the quantum dot, respectively. 
$\mu_i = \pm V/2$ is the chemical potential shift applied on the lead $i$ 
($V$ will denote the bias voltage throughout this paper), 
while the dot level is at zero chemical potential. 
Here, $b_{r}$ are the boson operators of the 
dissipative bath with an Ohmic type spectral density\cite{lehur2}: $\mathit{J}(\omega) = \sum_{r} \lambda_{r}^2 
\delta(\omega-\omega_{r}) = \alpha \omega$. 
Note that usually we introduce a cutoff via a $\exp(-\omega/\omega_c)$ 
function in $\mathit{J}(\omega)$; here, we assume that 
$\omega_c$ is a large energy scale comparable to the energy 
bandwidth of the reservoir leads.  
To simplify the discussion, we assume 
that the electron spins have been polarized 
through the application of a strong magnetic field. 
Note that our model can be realized experimentally 
in a quantum dot coupled 
to resistive environment as shown in Ref.~\onlinecite{finkelstein}. 

In this section, we briefly summarize the behavior of our  
model system at equilibrium which means in the absence of 
a finite bias voltage 
($V\!=\!0$).  
A dissipative resonant-level systems in equilibrium 
coupled to several leads 
maps onto the anisotropic one-channel Kondo 
model\cite{lehur2,Markus,matveev} 
where the dimensionless transverse Kondo coupling $g_{\perp}^{(e)}$  
is proportional to the hopping $t$ between the level and the leads  
and the longitudinal coupling 
$g_{z}^{(e)}\propto 1-\sqrt{\alpha }$ 
(the exact prefactors are given in Refs. 
\onlinecite{lehur2,Markus,matveev}; 
see also Sec. II B and Appendix A). Here, the superscript $(e)$ in 
$g_{\perp / z}^{(e)}$ refers to the equilibrium 
couplings. The model exhibits a Kosterlitz-Thouless (KT) QPT from a delocalized 
(Kondo screened) phase for $g_{\perp}^{(e)}+g_{z}^{(e)}>0$, 
with a large conductance, $G\approx e^2/h$, 
to a localized (local moment) phase for $g_{\perp}^{(e)}+g_{z}^{(e)}\leq 0$, 
with a small conductance, as the dissipation strength is increased 
(see Fig. 1). For $g_{\perp}^{(e)}\rightarrow 0$, the KT transition 
occurs at $\alpha _{c}=1$. 
As $\alpha\to\alpha_c$, the Kondo temperature $T_{K}$ 
obeys\cite{lehur1}: $\ln T_{K}\propto 1/(\alpha-\alpha_c)$. 
Note that here we assume our resonant level system exhibits the 
particle-hole (p-h) symmetry; namely, the resonant-level 
energy $\epsilon_d$ is set to be 
zero ($\epsilon_d=0$). However, in a more general resonant-level model 
where p-h symmetry is abscent, an additional term of the form 
$\epsilon_d d^{\dagger}d$ is present in the Hamiltonian Eq. (\ref{RLM}). 
In terms of its equivalent Kondo model, this p-h symmetry breaking 
term plays the role as an effective local magnetic field 
$B_z\propto \epsilon_d$ acting 
on the impurity spin in the Kondo model\cite{matveev}, 
which needs more involved treatments and exceeds the 
scope of a simple and generic model system considered in the present work.

In equilibrium, the dimensionless scaling functions 
$g_{\perp }^{(e)}(T)$ and $g_{z}^{(e)}(T)$ at the transition are 
obtained via the renormalization-group (RG) equations of the anisotropic 
Kondo model: 
\begin{equation}
g_{\perp ,cr}^{(e)}(T)=-g_{z,cr}^{(e)}(T)=(2\ln \left( {\mathcal{D}}/{T}\right) )^{-1}, 
\end{equation}
where $\mathcal{D}=D_{0}e^{1/(2g_{\perp })}$, with $D_{0}$ being the 
ultraviolet cutoff. Having 
in mind a quantum dot at resonance, $D_{0}=\min 
(\delta \epsilon ,\omega _{c})$, with $\delta \epsilon $ being the level 
spacing on the dot and $\omega _{c}$ the cut-off of the bosonic bath; 
$D_{0}$ is of the order of a few Kelvins. At low temperatures $T\ll D_0$, 
the conductance drops abruptly with decreasing temperatures\cite{zarand}: 
\begin{equation} 
G_{eq}(\alpha _{c},T\ll D_0)\propto \left[ g_{\perp ,cr}^{(e)}(T)\right] 
^{2}\propto \frac{1}{\ln ^{2}(T/\mathcal{D})}.  \label{GTeq}
\end{equation}

Below, we analyze the non-equilibrium $(V\neq 0$) transport of our model 
system at the KT transition and in the localized phase in the double-barrier 
resonant tunneling regime where the dissipative 
resonant level couples symmetrically 
to the two leads ($t_1=t_2=t$). Note, however, that 
when the dissipative resonant level couples asymmetrically 
to the leads $t_1\neq t_2$, as has been observed experimentally in 
Ref. ~\onlinecite{finkelstein},  
the system reaches the single-barrier tunneling regime, 
leading to Luttinger liquid behavior in conductance with power-law dependence 
in bias voltage.\\

For the sake of convenience, 
we set the following units throughout 
the rest of the paper: $e=\hbar=D_0=k_B=1$, and the temperature $T$ is in 
unit of $D_0=1$.

\subsection{\bf Useful Mappings}

Our generic model Hamiltonian Eq. (\ref{RLM}) 
in fact can be mapped onto various 
related model systems as we shall discuss below, 
including the anisotropic Kondo model (class (i)), the 
class (ii) and (iii) systems mentioned above. Here, we will address the 
non-equilibrium transport through a dissipative resonant level based on 
one of the equivalent models: the two-lead anisotropic Kondo model. 
 The mappings for the three classes of models discussed below will 
be derived in an analogous way. The general scheme of these mappings 
is via bosonization followed by re-fermionization 
(or in the opposite order)\cite{giamarchi,gogolin}.

\subsubsection{ {\bf Mapping the dissipative resonant level model onto the  
anisotropic  Kondo model}}

First, we envision a non-equilibrium mapping revealing that the leads are 
controlled by distinct chemical potentials. Through similar bosonization 
and 
refermionization procedures as in equilibrium, 
our model is mapped onto an 
anisotropic Kondo model\cite{lehur1,lehur2,Markus,matveev} 
with the effective (Fermi-liquid) left ($L$) 
and right lead ($R$)\cite{mapping} (see Appendix A for details):
\begin{eqnarray}
{H}_{K} &=&\sum_{k,\gamma =L,R,\sigma =\uparrow ,\downarrow }[\epsilon
_{k}-\mu _{\gamma }]c_{k\gamma \sigma }^{\dagger }c_{k\gamma \sigma } \nonumber \\
&+&(J_{\perp }^{(1)}s_{LR}^{+}S^{-}+J_{\perp
}^{(2)}s_{RL}^{+}S^{-}+h.c.)\nonumber \\
&+&\sum_{\gamma =L,R}J_{z}s_{\gamma \gamma
}^{z}S^{z},  \nonumber \\
\label{Hkondo}
\end{eqnarray}
where $c_{kL(R)\sigma }^{\dagger }$ is the electron operator of the 
effective lead $L(R)$, with $\sigma$ the spin quantum 
number, $\gamma=L, R$ is the index for the effective non-interacting 
fermionic leads, 
$S^{+}=d^{\dagger }$, $S^{-}=d$, and $S^{z}=Q-1/2$ where 
$Q=d^{\dagger }d$ describes the charge occupancy of the level. Additionally, 
$s_{\gamma \beta }^{\pm }=\sum_{\alpha ,\delta ,k,k^{\prime }}\frac{1}{2}c_{k\gamma 
\alpha }^{\dagger }\mathbf{\sigma }_{\alpha \delta }^{\pm }c_{k^{\prime 
}\beta \delta }$ are the spin-flip operators between the effective leads 
$\gamma $ and $\beta $, $J_{\perp }^{(1),(2)} \propto {t_{1,2}}$ embody the 
transverse Kondo couplings, $J_{z} \propto (1-{1}/\sqrt{2\alpha ^{\ast }})$, 
and $\mu _{\gamma }=\pm \frac{V}{2}\sqrt{1/(2\alpha ^{\ast })}$. 
It should be noted that this mapping is exact near the phase 
transition where $\alpha \rightarrow 1$ or 
$\alpha ^{\ast }\equiv \frac{1}{1+\alpha} \rightarrow 1/2$, 
and thus $\mu _{\gamma }=\pm V/2$. Note that the above 
mapping takes a spinless dissipative resonant level model with spinless 
fermionic baths $c_{\alpha=1,2}$ to the anisotropic Kondo model with a 
``spinful'' quantum dot (with spin operator given by $S^{+,-,z}$)  
and ``spinful'' conduction electron leads $\tilde{c}_{\gamma =L,R}^{\sigma}$. 
The appearance of the ``pseudo-spin'' degrees of freedom 
 in the effective Kondo model can be understood 
in terms of the "charge Kondo" effect: 
the the tunneling between a resonant level (which can be 
represented by a ``qubit'' or a ``spin'') and the 
spin-polarized leads plays an equivalent role as 
the ``pseudo-spin'' flips between spin of a quantum dot 
and that of the conduction electrons; and the coupling of 
the charge of the resonant level to the bosonic environment 
acts as the Ising coupling between z-components of the 
pseudo-spins on the dot and in the effective leads\cite{lehur1,lehur2}. 
Meanwhile, as mentioned above, when the resonant-level model shows 
p-h asymmetry, an additional term $\epsilon_d d^{\dagger}d$ appears in the 
Hamiltonian, which is equivalent to a local magnetic field acting on the 
impurity spin $\epsilon_d d^{\dagger}d \rightarrow B_z S_z $ 
via the identification: 
$d^{\dagger}=S^+$, $d=S^-$, and $d^{\dagger} d-1/2 = S_z$. 
For simplicity, we do not intend to study further this p-h asymmetry term 
and focus mainly on the effective Kondo model in the absence of 
magnetic field. Note also that the mapping has been derived earlier in 
Ref. \onlinecite{chung1} and is well-known at equilibrium 
(Ref. \onlinecite{lehur2}). In Appendix A, we will 
provide more details regarding the different theoretical steps, 
in particular with a finite bias voltage.

\subsubsection{{\bf Mapping for a resonant 
level coupled to a FQHE }}

Our analysis for the non-equilibrium transport of 
a dissipative resonant level 
model is applicable for describing a resonant level 
quantum dot coupled to two chiral Luttinger liquid leads, 
which is relevant for describing quasiparticle tunneling 
between two  
Fractional Quantum Hall Edge (FQHE) states\cite{FQHE} 
(the class (ii) model mentioned above). 
In the absence of bias voltage, this case has been studied 
in Refs. \onlinecite{lehur2,matveev}. 
Via the standard bosonization, 
\begin{equation}
c_{\alpha}(0) = \frac{1}{\sqrt{2\pi a}} F_{\alpha}  
e^{{\it i}\frac{\varphi_{\alpha}(0)}{K}}, 
\label{bosonization-LL}
\end{equation}
the Hamiltonian 
of such system can be written as\cite{FQHE,lehur1,lehur2,matveev} 
(see Appendix A.):
\begin{equation}
H_{FQHE} = H_{chiral} + H_{t} + H_{\mu}, 
\label{dot-luttinger}
\end{equation}
where the lead term $H_{chial}$ describes two chiral Luttinger liquid 
leads with lead index $\alpha=1,2$, $H_t$ denotes the tunneling term and the 
bias voltage term $H_{\mu}$ is given respectively by: 
\begin{eqnarray}
H_{chiral} &=& \frac{1}{2}\int_{-\infty}^{+\infty} 
\sum_{\alpha=1,2} \left(\frac{d\varphi_{\alpha}}{dx}\right)^2  dx,\nonumber \\
H_t &=& t_1 e^{{\it i}\varphi_1/\sqrt{K}} d + t_2 e^{{\it i}\varphi_2/\sqrt{K}} d + h.c.
\nonumber \\
H_{\mu} &=& -\frac{V}{2}\frac{1}{\sqrt{K}} (\partial \varphi_{1} - \partial \varphi_{2}),\nonumber \\
\label{H-LL}
\end{eqnarray}
where the boson field $\varphi_{\alpha=1,2}$ denotes the chiral Luttinger 
liquid in lead $\alpha$, the tunneling between lead and the resonant level 
is given by $t_{\alpha}$, $V$ is the bias voltage, and 
$K$ refers to the Luttinger parameter. Here, we set $2\pi a =1$ throughout 
the paper with $a$ being the lattice constant. 
Through the similar refermionization, we arrive at the effective Kondo model 
as shown in Eq. (~\ref{Hkondo}) with the bare Kondo couplings 
$J_{\perp}^{(1),(2)} = t 
e^{{\it i} (\sqrt{2} - \frac{1}{\sqrt{K}}) \varphi_{2,1}}$, 
$J_{z} = 1 - 1/\sqrt{ 2 K}$. The non-equilibrium RG scaling 
equations for $H_{FQHE}$ have the same form as in Eq. (\ref{RG2}).\\

\subsubsection{{\bf Mapping for a dissipative resonant level coupled to 
interacting leads}}

So far, we consider here just  
a dissipative resonant single-level coupled to 
two non-interacting leads. Nevertheless, the mapping can be 
straigthforwardly generalized to 
the same system but with a spinless quantum dot 
which contains many energy levels. In this case, the effective Luttinger liquid 
parameter $K'$ is modified as: 
$\frac{1}{K'} = \frac{1}{K} +1$ (see Eq. (~\ref{Kprime}) in Appendix A). 
More generally, 
the mapping can be further 
generalized to the system of a many-level (single-level) 
spinless quantum dot with Ohmic dissipation coupled to 
two chiral Luttinger liquid leads with Luttinger parameter $K$, 
giving rise to the effective Luttinger 
liquid parameter $\tilde{K}$ defined as (see Eq. (~\ref{Ktilde}) in Appendix A):
\begin{equation}
\tilde{K}= \frac{1}{K} + 1 + K_b
\end{equation}
 for a many-level spinless   
quantum dot and 
\begin{equation}
\frac{1}{\tilde{K}} = \frac{1}{K} + K_b
\label{K-small}
\end{equation}
 for a spinless quantum dot with 
a single resonant level. Details of the mapping is given in Appendix A.

\subsection{Average current}

We may compute the non-equilibrium 
current operator in the effective models through the 
mappings. We will first compute the current operator within the effective 
anisotropic Kondo model as it is the main focus of this paper.  
From the mapping described in Sec. II. B {\it 1.}, 
we can establish  the invariance of the net charge on the resonant level 
upon the mapping: $N_1-N_2 = N_{L}-N_{R}$, where 
$N_i=\sum_{ki} c^{\dagger}_{ki} c_{ki}$ represents the charge in lead $i=1,2$, whereas $N_{\gamma}= \sum_{k} c^{\dagger}_{k\gamma\sigma} c_{k\gamma\sigma} 
$ represents the charge in the effective lead $\gamma=L,R$. 
This allows us to check that the averaged currents within the 
Keldysh formalism\cite{keldysh} are the same in the original and in the effective Kondo 
model (see Appendix B for details): 
\begin{eqnarray} 
I 
&=& {\it i} [Q_{L} - Q_{R}, H_K] \nonumber \\ 
&=& {\it i} J_{\perp}^{(1)} (s_{LR}^{-} S^{+} - s_{RL}^{+} S^{-}) - 
   (1\rightarrow 2, L\rightarrow R).\nonumber \\
\label{currentI}
\end{eqnarray} 
Thus, the current $I$ can be computed from the Kondo model due to 
the invariance of the average current upon the mapping mentioned above. 
Note that through the various mappings mentioned above, 
it is straightforward to see that 
the current operator for other related models--resonant level coupled 
to FQHE leads and dissipative resonant level (both small and large in size) 
coupled to interacting Luttinger liquid leads--take exactly the same form 
as shown in Eq. (\ref{currentI}).

\section{\bf Non-equilibrium RG approach}

\subsection{RG equations}

Now, we employ the non-equilibrium RG approach to the effective 
Kondo model\cite{noneqRG} in Eq. (\ref{Hkondo}). 
In this approach, the Anderson's 
poor-man scaling equations are generalized to 
non-equilibrium RG equations by including the frequency dependence of 
the Kondo couplings and the decoherence due to the steady-state current at 
finite bias voltage\cite{noneqRG}. For the sake of simplicity, 
we assume that the resonant level (quantum dot) is symmetrically 
coupled to the right and to the left lead, $t_{1}=t_{2}$ 
(or $J_{\perp}^{(1)}=J_{\perp}^{(2)}\equiv J_{\perp}$). 
We will discuss in Appendix C the more general 
case with $t_1\neq t_2$. 
The dimensionless Kondo couplings as a function of frequency $\omega$ exhibit 
an extra symmetry due to the particle-hole symmetry of the effective 
Kondo model: $g_{\perp (z)}(\omega 
)=g_{\perp (z)}(-\omega )$ where $g_{\perp (z)}(D_0)=N(0)J_{\perp (z)}$ 
is the initial value, with $N(0) 
$ being the density of states per spin of the conduction electrons. 
Here, we suppress the upper script symbol $^{(e)}$ in the Kondo couplings 
since we will now focus on the non equilibrium case V not zero. 
We obtain\cite{noneqRG}: 
\begin{eqnarray}
\frac{\partial g_{z}(\omega )}{\partial \ln D} &=&-\sum_{\beta =-1,1}\left[ 
g_{\perp }\left( \frac{\beta V}{2}\right) \right] ^{2}\Theta _{\omega +
\frac{\beta V}{2}},   \nonumber \\
\frac{\partial g_{\perp }(\omega )}{\partial \ln D} &=&-\!\!\!\sum_{\beta 
=-1,1}\!\!\!g_{\perp }\left( \frac{\beta V}{2}\right) g_{z}\left( \frac{\beta V}{2}\right) \Theta _{\omega +\frac{\beta V}{2}},  
\label{gpergz}
\end{eqnarray}
where $\Theta _{\omega }=\Theta (D-|\omega +\mathit{i}\Gamma |)$, $D<D_{0}$ 
is the running cutoff. Here, $\Gamma$ 
is the decoherence (dephasing) rate at 
finite bias which cuts off the RG flow\cite{noneqRG}. 
In the Kondo model, $\Gamma$ corresponds to the relaxation rate due to 
spin flip processes (which are charge flips in the original model),  
defined as the 
broadening $\Gamma =\Gamma_s$ of the dynamical transverse spin 
susceptibility $\chi^{\perp}(\omega)$ in the effective Kondo 
model\cite{decoherence}:
\begin{equation}
\chi^{\perp}(\omega) = 
\chi_0 \frac{{\it i}\Gamma_s}{\omega + {\it i}\Gamma_s} 
\end{equation} 
with $\chi^{\perp}(\omega)$ being the time Fourier transform of 
the spin susceptibility $\chi^{\perp}(t)={\it i}\theta(t)\langle 
[S^{-}(t),S^{+}(0)]\rangle = 
{\it i}\theta(t)\langle [f^{\dagger}_{\downarrow}(t) f_{\uparrow}(t), 
f^{\dagger}_{\uparrow}(0) f_{\downarrow}(0)]\rangle$, and $\chi_0$ being 
$\chi^{\perp}(\omega=0)$. Here, we take the pesudo-fermion representation 
of the spin operators 
$S^{+,-,z} = \frac{1}{2} f_{\alpha}\mathcal{\sigma}^{+,-,z}_{\alpha\beta}f_{\beta}$ 
with $f_{\sigma=\uparrow,\downarrow}$ being the pesudo-fermion 
operator and $\mathcal{\sigma}^{+,-,z}$ being the Pauli 
matrices\cite{noneqRG}.\\ 

In the original model the decoherence rate 
$\Gamma$ corresponds to the charge flip rates, 
defined as the broadening $\Gamma_d$ of the 
resonant-level ($d-$electron) Green's function (or equivalently the 
imaginary part of the resonant-level self-energy $Im(\Sigma_d(\omega))$):  
$\Gamma=\Gamma_d = Im(\Sigma_d(\omega))$ 
where the self-energy $\Sigma_d(\omega)$ of the 
$d-$electron Green's function is defined via:  
$1/G_d(\omega) \propto \omega+\epsilon_d + Re(\Sigma_d(\omega)) 
+ {\it i} Im(\Sigma_d(\omega))$ with $G_d(\omega)$ being the 
Fourier transform of the resonant-level Green's function 
$G_d(t) ={\it i} \theta(t) \langle[d(t), d^{\dagger}(0)]\rangle$. 
 These two definitions for $\Gamma$ agree with each other 
with the proper identification: $d=S^{-}$, $d^{\dagger} = S^{+}$.

\begin{figure}[t]
\begin{center}
\includegraphics[width=8.0cm]{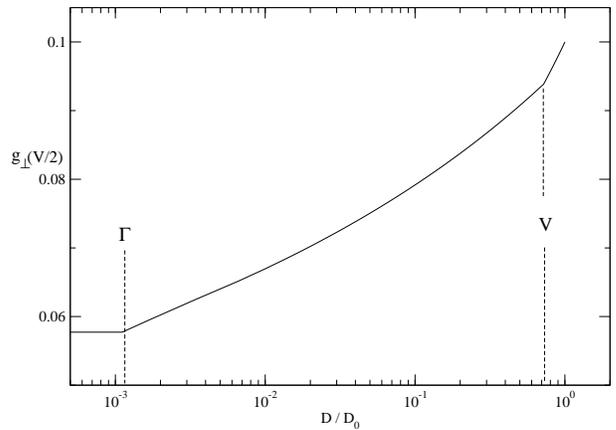}
\end{center}
\par
\vspace{-0.5cm} 
\caption{RG flow of $g_{\perp,cr}(V/2)$ at the 
transition as a function of bandwidth cutoff $D$ (in unit of $D_0$); 
the bare couplings are $g_{\perp}=-g_z = 0.1$ (in unit of $D_0$). 
We have set $V = 0.72$ (in unit of $D_0$). The 
decoherence rate $\Gamma$ is around $0.00117D_0$.}
\label{RGflow}
\end{figure}

Note that these RG equations in the present context were already 
discussed in the short Ref.~\onlinecite{chung1}, 
but now we will elaborate the methodology.
The configurations of 
the system out of equilibrium are not true eigenstates, but acquire a finite 
lifetime. The spectral function of the fermion on the level is peaked 
at $\omega =\pm V/2$, and therefore we have $g_{\perp (z)}(\omega )\approx 
g_{\perp (z)}(\pm V/2)$ on the right hand side of Eq. (\ref{gpergz}). 
Other Kondo couplings are not generated. 
From Ref.~\onlinecite{noneqRG} via the Fermi's golden rule of the 
spin-flip rates $\Gamma$ in the Kondo model, we identify: 
\begin{eqnarray}  \label{gammaV} 
\Gamma = \frac{\pi}{4} \sum_{\gamma,\gamma^{\prime},\sigma} \int \!\!
d\omega \Bigl[ n_{\sigma} g_{z}^2(\omega)
f_{\omega-\mu_{\gamma}}(1-f_{\omega-\mu_{\gamma ^{\prime}}}) && \\
+ n_{\sigma} g_{\perp}^2(\omega)
f_{\omega-\mu_{\gamma}}(1-f_{\omega-\mu_{\gamma^{\prime}}})\Bigr], &&
\nonumber
\end{eqnarray}
where $f_{\omega}$ is the Fermi function. Here, $\gamma=\gamma^{\prime}$ for 
the $g_z^2(\omega)$ terms while $\gamma \neq \gamma^{\prime}$ for the 
$g_{\perp}^2(\omega)$ terms with $\gamma$, $\gamma^{\prime}$ being $L$ or $R$. 
We have introduced the occupation numbers $n_\sigma$ for up and down 
spins satisfying $n_{\uparrow} +n_{\downarrow}=1$ and $S_z=(n_{ 
\uparrow}-n_{\downarrow})/2$. In the delocalized phase, we get $n_{\uparrow} = 
n_{\downarrow} = 1/2$, in agreement with the quantum Boltzmann equation 
\cite{noneqRG}. At the KT transition, we can use that 
$g_{\perp}(\omega)=-g_z(\omega)$ from the symmetry of the Kondo model 
and that $\sum_{\sigma} n_{\sigma} =1$. Finally in the localized phase, we have 
$g_{\perp}\le -g_z$, and $n_{\sigma}$ satisfies $|S_z| 
\rightarrow 1/2$ (see Refs.~\onlinecite{lehur1,lehur2,Markus,matveev}), which remains true at a finite bias voltage.
\begin{figure}[t]
\begin{center}
\includegraphics[width=8.0cm]{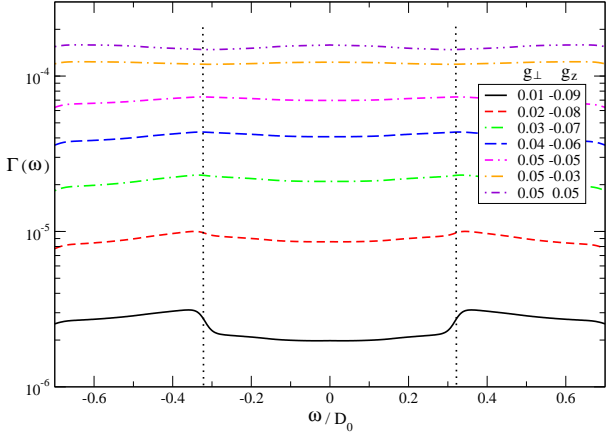}
\end{center}
\par
\vspace{-0.5cm} 
\caption{(Color online) $\Gamma(\protect\omega)$ at $T=0$ versus $\omega$ 
across the KT transition. 
The bias voltage 
is fixed at $V=0.32 D_0$. $\Gamma(\omega)$ develops a peak (dip) 
at $\omega=0$ in the delocalized (localized) phase, respectively. 
At $\omega\approx \pm V$ (vertical dotted lines), 
$\Gamma(\omega)$ shows peaks (for localized phase) 
or dips (for delocalized phase). Note that $\Gamma(\omega)$ 
 weakly depends on $\omega$ for $|\omega|<V$, 
$\Gamma(|\omega|<V)\approx \Gamma(\omega=0)$.}
\label{gamma_omega}
\end{figure}

\subsection{Solutions to RG equations}

Following the scheme of Ref. \onlinecite{noneqRG}, we solve Eqs. (\ref{gpergz})
and (\ref{gammaV}) self-consistently. First, we compute $g_{\perp(z)}(\omega
=\pm V/2)$ for a given cutoff $D$. We then substitute the solutions back 
into the RG equations to get the general solutions for $g_{\perp(z)}(\omega)$ 
at finite $V$, and finally extract the solutions in the limit $D\to 0$. When 
the cutoff $D$ is lowered, the RG flows are not cutoff by $V$ but they 
continue to flow for $\Gamma < D < V$ until they are stopped for 
$D\le \Gamma$.\\ 

In Fig.\ref{RGflow} we show a typical RG flow of 
$g_{\perp}(V/2)$ at the KT transition 
as a function of bandwidth $D$ with the analytical 
approximation: $g_{\perp}(V/2)\approx \frac{1}{2\ln\frac{\mathcal{D}}{D}}$ 
for $D>V$, $g_{\perp}(V/2)\approx \frac{1}{\ln\frac{\mathcal{D}^2}{DV}}$ 
for $\Gamma<D<V$, and $g_{\perp}(V/2)\approx 
\frac{1}{\ln\frac{\mathcal{D}^2}{V\Gamma}}$ 
for $D<\Gamma$. Here, 
$\mathcal{D}=D_{0}e^{1/(2g_{\perp })}$, with $D_{0}$ being the 
ultraviolet cutoff, and $D$ is the running cutoff scale set by the RG 
scaling equations for $g_{\perp/z}$. 
This clearly shows that the RG flow of $g_{\perp}(V/2)$ 
is stopped at $\Gamma$, a much lower energy scale than $V$.\\ 

\begin{figure}[t]
\begin{center}
\includegraphics[width=8.0cm]{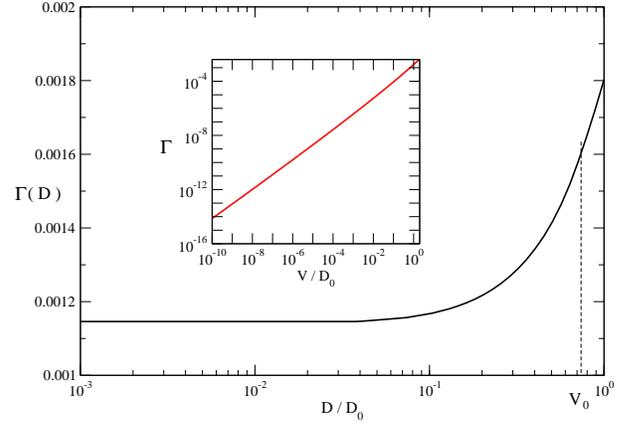}
\end{center}
\par
\vspace{-0.5cm} 
\caption{(Color online) RG flow of 
$\Gamma(D)$ at $T=0$ versus $D$ (in unit of $D_0$) 
at a fixed bias voltage $V_0 = 0.72 D_0$ (vertical dashed line) 
at the KT transition 
with bare Kondo couplings $g_{\perp}=0.1=-g_z$. Under RG, 
$\Gamma$ approaches a constant value as $D\rightarrow 0$:  
$\Gamma (D\rightarrow 0) \approx 0.00117D_0$.  
Inset: $\Gamma$ as a function of $V$ for the same bare 
Kondo couplings.}
\label{Gamma-D}
\end{figure}

Note that the charge (or pseudospin) decoherence rate $\Gamma$ is  
a function of frequency, $\Gamma(\omega)$ in the more general and rigorous 
Functional 
Renormalization Group (FRG) framework\cite{FRG}. Here, 
$\Gamma=\Gamma(\omega=0)$ within FRG. 
Nevertheless, we find $\Gamma(|\omega|\le V)$ at $T=0$ depends 
weakly on $\omega$ and 
can be well approximated by its value 
at $\omega=0$, $\Gamma(T=0,\omega) \approx \Gamma(T=0,\omega=0)$ 
(see Fig.~\ref{gamma_omega}). 
We have checked that the non-equilibrium 
current ${\it I}(V,T=0)$ and 
conductance $G(V,T=0)$ obtained from this approximation 
($\Gamma(T=0,\omega) \approx \Gamma(T=0,\omega=0)$) agrees very well with that 
from the more rigorous FRG approach based on the frequency-dependent 
decoherence rate $\Gamma(T=0,\omega)$ (see Eq. (~\ref{gamma}) below) 
as a consequence of the 
fact that the current and conductance are 
integrated quantities over the frequencies, and they are insensitive to 
weak frequency-dependence of $\Gamma$.  
In Fig.~\ref{Gamma-D} we show the RG flow of the 
decoherence rate $\Gamma(\omega=0)$ 
as a function of $D$, using the same parameter as in Fig. 1. 
One observes that $\Gamma$ tends to a finite value as $D\rightarrow 0$. 
The inset shows $\Gamma$ as a function of $V$ (see also section VI). 
Note that, unlike the equilibrium RG at finite temperatures 
where RG flows are cutoff by temperature $T$,  
 here in non-equilibrium the RG flows will be cutoff by the 
decoherence rate $\Gamma$, 
an energy scale typically much higher than $T$, but much lower than $V$, 
$T\ll \Gamma \ll V$. Moreover, $\Gamma(V)$ is a non-linear function in $V$. 
(For example, at the KT transition, $\Gamma_{KT}(V)
\propto V/ [\ln(\frac{{\mathcal D}}{V})]^2$. 
). The unconventional properties of $\Gamma(V)$ lead to 
a non-equilibrium conductance ($G(V,T=0)$) distinct 
from that in equilibrium ($G(T,V=0)$) 
near the KT transition\cite{chung1}. In contrast, the 
equilibrium RG will lead to approximately frequency independent 
couplings, (or ``flat'' functions 
$g_{\perp}(\omega)\approx g_{\perp,z}(\omega=0)$).\\

Notice that the mapping mentioned above 
works near the KT transition, 
$\alpha^{\ast}\equiv \frac{1}{1+\alpha}\rightarrow 1/2$. 
However, for a general case deep in the localized phase, 
the effective Kondo couplings aquire an additional phase 
$J_{\perp}^{(1),(2)} \propto t_{1,2} e^{{\it i} (\sqrt{2} - \frac{1}{\sqrt{K}}) 
\tilde{\phi}_{s;2,1}}$ where the more general form of $J_{\perp}^{(1),(2)}$ 
and its phase $\tilde{\phi}_{s;2,1}$ are derived 
and defined in Eq.~(\ref{tilde-H-t}) 
of Appendix A.. This results in a nonzero bare scaling 
dimension\cite{giamarchi} 
for $J_{\perp}^{(1),(2)}$, $[J_{\perp}^{(1),(2)}] 
= \frac{1}{2}(\sqrt{2} - \frac{1}{\sqrt{K}})^2 = 1-\sqrt{\frac{2}{K}} 
+ \frac{1}{2K}$. 
This slightly modifies the non-equilibrium RG scaling equations 
to the following form: 
\begin{eqnarray}
\frac{\partial g_{z}(\omega )}{\partial \ln D} &=&-\sum_{\beta =-1,1}\left[ 
g_{\perp }\left( \frac{\beta V}{2}\right) \right] ^{2}\Theta _{\omega + 
\frac{\beta V}{2}} \nonumber \\
\frac{\partial g_{\perp }(\omega )}{\partial \ln D} &=&-\!\!\!\sum_{\beta 
=-1,1}\!\!\! [ 
\frac{1}{2}(1-\sqrt{\frac{2}{K}} + \frac{1}{2K}) g_{\perp }\left( \frac{\beta V}{2}\right) 
\nonumber \\
&+& 
g_{\perp }\left( \frac{\beta V}{2}\right) g_{z}\left( \frac{\beta V}{2}\right)
 ]
\Theta _{\omega +\frac{\beta V}{2}} 
\label{RG2}
\end{eqnarray}
where the linear term $\frac{1}{2}(1-\sqrt{\frac{2}{K}} + \frac{1}{2K}) g_{\perp }\left( \frac{\beta V}{2}\right)$ in Eq.~(\ref{RG2}) for $g_{\perp}(\omega)$ 
comes from the bare scaling dimension of $J_{\perp}^{(1),(2)}$ terms mentioned 
above, and it vanishes in the limit of $K\to 1/2$, as expected. 
In fact, this term applies to the three models (case (i), (ii) and (iii)) 
through the mappings. 
Note that the above scaling equations may be cast in the same form as 
in Eq. ~(\ref{gpergz}) through redefinition of the coupling $g_z$: 
$g_z\rightarrow \bar{g}_z=
g_z +\frac{1}{2}(1-\sqrt{\frac{2}{K}} + \frac{1}{2K})$. All 
the previous results remain valid upon the above shift of $g_z$.\\
\begin{figure}[t]
\begin{center}
\includegraphics[width=7.5cm]{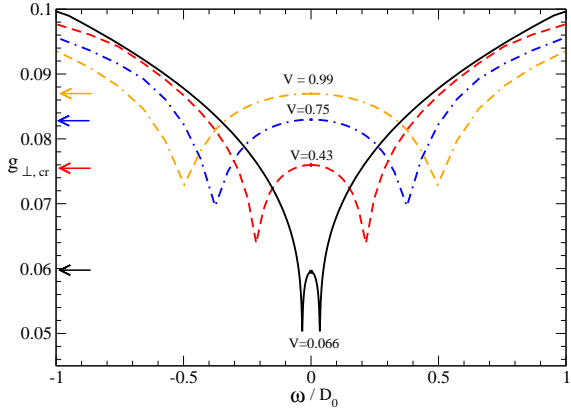}
\end{center}
\par
\vspace{-0.5cm} 
\caption{$g_{\perp,cr}(\protect\omega)= -g_{z,cr}(\protect\omega)$ at the 
transition at various bias voltages $V$ (in unit of $D_0$); 
the bare couplings are $g_{\perp}=-g_z = 0.1$ (in unit of $D_0$). The 
arrows give the values of $g_{\perp}(\protect\omega = 0)$ at these bias 
voltages. }
\label{gpergzfig}
\end{figure}

\section{\bf Non-equilibrium conductance}

In the section, we present our results for non-equilibrium conductance. 
All explicit results will be obtained for the KT transition point and 
the localized phase, but not for the delocalized phase.

\subsection{\bf Non-equilibrium conductance at the KT transition}

At the KT transition, we both numerically and analytically solve Eqs. 
(\ref{gpergz}) and (\ref{gammaV}) (in the limit of $D\to 0$). 
In particular, the approximated analytical solution 
within the approximation 
$\Theta _{V }\approx \Theta (D-V)$ due to $\Gamma \ll V$ is obtained:   
\begin{eqnarray}  \label{gsolution} 
g_{\perp,cr}(\omega) &\approx& 
\sum_{\beta} \Theta (|\omega-\beta V/2|-V) \frac{1}{
4 \ln\left[\frac{\mathcal{D}}{|\omega-\beta V/2|}\right]} \\
&+& \Theta (V-|\omega-\beta V/2|) \times  \nonumber \\
&& \left[ \frac{1}{\ln[\mathcal{D}^{2}/ V \max(|\omega-\beta V/2|,\Gamma)]}
- \frac{1}{4 \ln\frac{\mathcal{D}}{V}}\right].  \nonumber
\end{eqnarray}
The solutions at the transition (denoted $g_{\perp,cr}$ and $g_{z,cr}$) are shown in Fig. \ref{gpergzfig}. Since $g_{\perp,cr}(\omega)$ 
decreases under the RG scheme, the effect of the 
decoherence leads to minima; the couplings are 
severely suppressed at the points $\omega=\pm \frac{V}{2}$. We 
also check that $g_{\perp,cr}(\omega) = - g_{z,cr}(\omega)$.\\

\begin{figure}
\includegraphics[width=7.5cm]{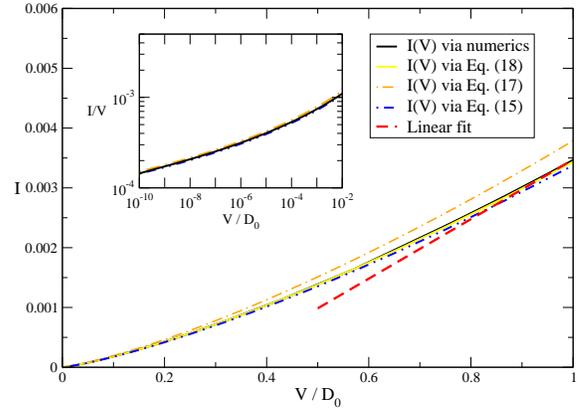}
\vspace{-0.2cm}
\caption{Non-equilibrium current at the localized-delocalized transition. The 
n\"aive approximate analytical expression 
Eq.~(\ref{IVapproxnaive}) fits well with the 
numerical result 
over several decades (from $V\approx 10^{-12}D_0$ to 
$V\approx 10^{-3}D_0$, see Inset). However, it starts to deviate 
from the numerical result at higher bias voltages.} 
\label{I}
\end{figure}

From the Keldysh calculation up to second order in the tunneling amplitudes, 
the current reads:
\begin{eqnarray}  \label{noneqI}
I = \frac{ \pi}{8} \int d\omega \Big[\sum_\sigma 4g_{\perp}(\omega)^2 
n_{\sigma}\times \\
f_{\omega-\mu_{L}} (1-f_{\omega-\mu_{R}})\Big] - (L \leftrightarrow R). 
\nonumber
\end{eqnarray}
At $T=0$, it simplifies as $I = \frac{\pi}{2} \int_{-V/2}^{V/2}
\! d\omega  g_{\perp}^{2}(\omega)$. Then, we numerically evaluate the 
non-equilibrium current. The differential 
conductance is obtained as $G(V) = dI/dV$. The 
$T=0$ results at the KT transition are shown in Fig.~\ref{I} and 
Fig.~\ref{Icr}.

First, it is instructive to compare the non-equilibrium current at the transition to the (na\"ive) approximation: 
\begin{eqnarray}
\label{IVapproxnaive}
I_{cr} &\approx & \frac{\pi V}{2} 
\left[g_{\perp,cr}(\omega = 0)\right]^2
\approx  \frac{\pi}{8 } 
\frac{V}{(\ln^2({T_D}/{V}))}.
\end{eqnarray}

As shown in Fig.~\ref{I}, our numerically obtained non-equilibrium 
current fits well with the above analytical approximation for $V<0.01D_0$. 
However, it starts to deviate from its numerically 
 obtained values for higher bias voltages $V > 0.01 D_0$. 
This deviation is due to the fact that 
the equilibrium form of the conductance at the transition is obtained 
by treating $g_{cr\perp}(\omega)$ a flat function within 
$-V/2 < \omega < V/2$: 
$g_{cr\perp}(\omega)\approx g_{cr\perp}(\omega=0) 
\approx g^{eq}_{cr\perp}(T\to V)$.
We have checked that the 
equilibrium coupling $g_{cr\perp}(\omega =0)$ indeed corresponds 
to $g_{\perp}^{eq}(T=V)$, therefore the transport recovers the 
expected equilibrium form for $V\to 0$. 
However, since $g_{\perp}(\omega)$ is not a flat 
function for $-V/2<\omega <V/2$ (it has two minima at $\omega = \pm V/2$), 
with increasing $V$ (say for $V \approx 0.01D_0$) 
the non-equilibrium current exhibits a distinct behavior due to 
the frequency dependence of the coupling.

 In fact, the more accurate approximate expression for the 
 non-equilibrium current at the transition is found to be:

\begin{figure}[t]
\begin{center}
\includegraphics[width=7.5cm]{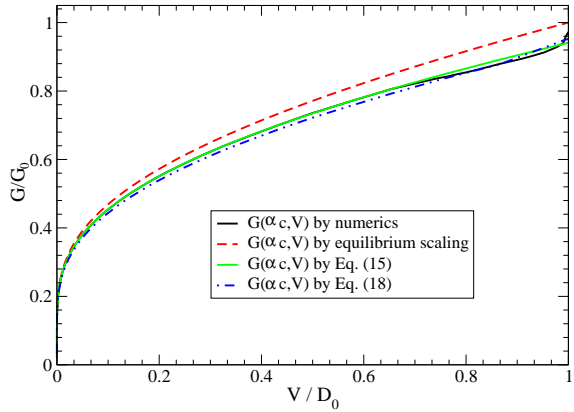}
\end{center}
\par
\vspace{-0.1cm}
\caption{Non-equilibrium conductance $G=dI/dV$ at the KT transition. $G_0$ is the 
equilibrium conductance at the transition for $T=D_0$: 
$G_0 = G_{eq}(\protect \alpha_c, T=D_0) = 0.005 \protect\pi$ 
with the bare couplings $g_{\perp} = -g_z = 0.1D_0$. }
\label{Icr}
\end{figure}
\begin{eqnarray}  
\label{IVapprox}
I(\alpha_c,V) &\approx & \frac{\pi V}{2} \left( \frac{\pi}{4} \left[
g_{\perp,cr}(\omega = 0)\right]^2 \right) \\ 
& +& \frac{\pi V}{2} \left( (1-\frac{\pi}{4}) \left[g_{\perp,cr}(\omega
= V/2)\right]^2 \right),  \nonumber
\label{Icr-app}
\end{eqnarray}
where 
\begin{eqnarray}
g_{\perp,cr}(\omega= V/2) &\approx& 1/\ln(\frac{{\mathcal{D}}^2}{\Gamma V})\\ \nonumber
g_{\perp,cr}(\omega = 0) &\approx& 2 \left(\frac{1}{\ln(2{\mathcal{D}}
^2/V^2)} - \frac{1}{4\ln({\mathcal{D}}/V)}\right).
\label{gcr-app}
\end{eqnarray}
\begin{figure}[h]
\begin{center}
\vspace{0.7cm}
\includegraphics[width=8.5cm]{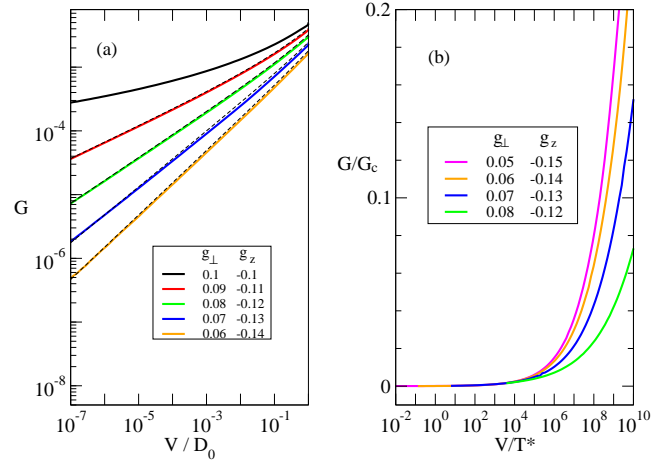}
\end{center}
\par
\vspace{-0.3cm}
\caption{
Conductance in the localized phase (in units of $\pi$). (a) $G(V)$ at low bias follows the equilibrium scaling 
(dashed lines). (b) The conductance $G(V)/G_c$ is a function of $V/{T}^{\ast}$ where we have defined $G_c=G(\protect\alpha_c, V)$ and 
${T}^{\ast}=D_0 e^{-\protect\pi/\protect\sqrt{g_z^2-g_{\perp}^2}}$.}
\label{GvA}
\end{figure}
 Here, we have treated $g_{\perp,cr}(\omega)^2$ within the interval 
$-V/2<\omega<V/2$ as a semi-ellipse.\\

As demonstrated in Fig.~\ref{Icr}, the conductance $G(V)$ obtained via 
the approximation in Eq.~(\ref{IVapprox}) fits very well with that obtained 
numerically over the whole range of $0<V<D_0$. In the low-bias $V\to 0$
(equilibrium) limit, since $g_{\perp,cr}(\omega = 0)\approx
g_{\perp,cr}^{(e)}(T=V)\ll 1$, we have $I(\alpha_c,V)\approx \frac{\pi V}{2}
\left( g_{\perp,cr}^{(e)}(T= V)\right)^2$; therefore the scaling of 
$G(\alpha_c, V)$ is reminiscent of the equilibrium expression in 
Eq. ~(\ref{GTeq}), $G(\alpha_c,V) \approx \frac{\pi}{2} \left(g_{
\perp,cr}^{(e)}(T=V)\right)^2 = \frac{\pi}{8}\frac{1}{\ln^2({\mathcal{D
}}/V)}$. This agreement between equilibrium and non-equilibrium conductance 
at low $V$ persists up to a crossover scale $V\approx 0.01D_0$ (determined for the parameters used 
in Fig. \ref{Icr}). At larger biases, the conductance shows a unique non-equilibrium profile; see Eq. (\ref{IVapprox}). 
We find an excellent agreement of the non-equilibrium conductance 
obtained by three 
different ways --- pure numerics, analytical solution Eq. (\ref{gsolution})
and the approximation in Eq. (\ref{IVapprox}).

For large bias voltages $V\rightarrow D_{0}$, since 
$g_{\perp,cr}(\omega )$ approaches its bare value $g_{\perp}$, the non-equilibrium conductance increases rapidly and reaches $G(\alpha _{c},V)\approx G_{0}=\frac{\pi }{
2}g_{\perp}^{2}$. Note that the non-equilibrium conductance is 
always smaller than the equilibrium one, $G(\alpha _{c},V)<G_{eq}(\alpha _{c},T=V)$, since $g_{\perp 
}(\omega =\pm V/2)<g_{\perp }(\omega =0)$. 
Additionally, in the delocalized phase for 
$V\gg T_{K}>0$, the RG flow of $g_{\perp}$ is suppressed by the 
decoherence rate, and $G\propto 1/\ln ^{2}(V/T_{K})$ (Ref.~\onlinecite{noneqRG}).\\

\subsection{\bf Non-equilibrium conductance in the localized phase}

In the localized phase, we first solve the equilibrium RG equations
of the effective Kondo model analytically, resulting in 
\begin{eqnarray}
 G_{loc}^{(e)}(T)&=& \frac{\pi}{
2} \left(g_{\perp,loc}^{(e)}(T) \right)^2\\ \nonumber
 g_{\perp,loc}^{(e)}(T) &=& \frac{2cg_{\perp} (c+|g_z|)}{(c+|g_z|)^2 - g_{\perp}^2
(\frac{T}{D_0})^{4c}} (\frac{T}{D_0})^{2c}\nonumber \\
\label{GT-loc}
\end{eqnarray}
  where $c = \sqrt{g_z^2 -g_{\perp}^2}$. We introduce the energy scale 
${T}^{\ast}=D_0 e^{-\pi/\sqrt{g_z^2-g_{\perp}^2}}$ (which vanishes at the KT transition) such that $g_{\perp,loc}^{(e)}(T)\propto ({T}/T^{\ast})^{2c}$  for $T\to 0$, leading to 
$G_{loc}^{(e)}(T)\propto ({T}/T^{\ast})^{4c}$.\\
\begin{figure}[h]
\begin{center}
\vspace{0.3cm} 
\includegraphics[width=8.25cm]{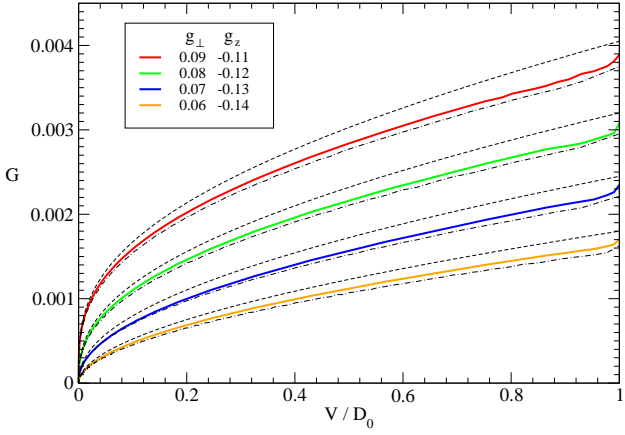}
\end{center}
\par
\vspace{-0.6cm}
\caption{
Conductance in the localized phase (in units of $\pi$). 
At large bias voltages $V$, the non-equilibrium 
conductance $G(V)$ (solid lines) is distinct from the equilibrium form 
(dashed lines). The dot-dashed lines stem from an analytical approximation 
via Eq. (\ref{IVapproxloc}).}
\label{GvB}
\end{figure}
At a finite bias, we first solve for the self-consistent 
non-equilibrium RG equations both analytically and numerically,  
resulting in:
\begin{eqnarray}
\label{gp-loc-app-a}
&g_{\perp,loc}&(\omega=V/2) \approx  
g_{\perp} + \frac{A}{2 c} [V^{2c}\sqrt{c^2+A^2 V^{4c}}\\ \nonumber
&-& \sqrt{A^2+c^2}] + 
\frac{B}{2c} [\Gamma^c \sqrt{c^2+B^2\Gamma^{2c}} \\ \nonumber
&-& V^c \sqrt{c^2+B^2V^{2c}}] \\ \nonumber 
&+& \frac{c}{2} \ln\left[ \frac{B\Gamma^c +\sqrt{c^2+
B^2\Gamma^{2c}}}{BV^{c} +\sqrt{c^2+B^2 V^{2c}}}\right ]\\ \nonumber
&+& \frac{c}{2} \ln\left[
\frac{A V^{2c} +\sqrt{c^2+A^2 V^{4c}}}{A +\sqrt{c^2+A^2}}\right ],
\end{eqnarray}
\begin{eqnarray}
\label{gp-loc-app-b}
&g_{\perp,loc}&(\omega=0) \approx  g_{\perp} + \frac{A}{2 c} 
[V^{2c}\sqrt{c^2+A^2 V^{4c}}\\ \nonumber
&-& \sqrt{A^2+c^2}]
+ \frac{B}{c} [ (\frac{V}{2})^c \sqrt{c^2+B^2(\frac{V}{2})^{2c}}\\ \nonumber
&-& V^c \sqrt{c^2+B^2 V^{2c}}]\\ \nonumber
&+& \frac{c}{2} \ln\left [\frac{B(\frac{V}{2})^c +
\sqrt{c^2+B^2 (\frac{V}{2})^{2c}}}{BV^{c} +\sqrt{c^2+B^2 V^{2c}} }\right ]
\\ \nonumber
&+&
\frac{c}{2} \ln\left [\frac{A V^{2c} +\sqrt{c^2+
A^2 V^{4c}}]}{[A +\sqrt{c^2+A^2} }\right],
\end{eqnarray} 
and similarily we get 
\begin{eqnarray}
\label{gz-loc-app-a}
&g_{z,loc}&(\omega=V/2) \approx g_{z} + \frac{A^2}{2c} [1-V^{4c}] \\ \nonumber
&+& \frac{B^2}{2c} V^{2c} [1-(\frac{\Gamma}{V})^{2c}],  
\end{eqnarray}
\begin{figure}[!tbp]
\begin{center}
\vspace{0.5cm}
\includegraphics[width=9cm]{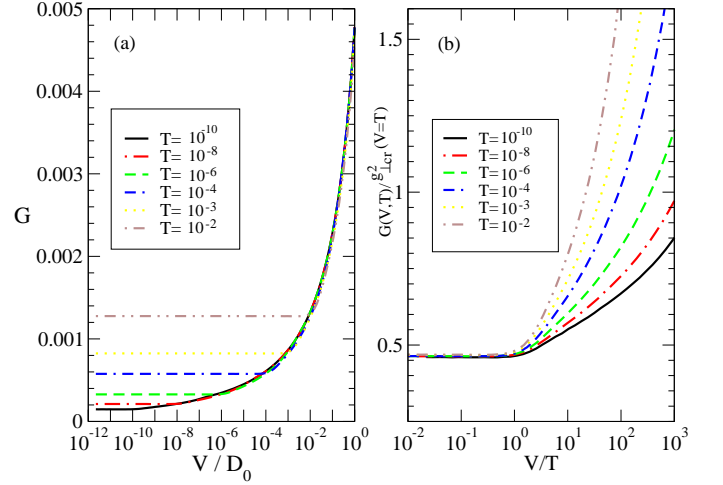}
\end{center}
\par
\vspace{-0.50cm} 
\caption{Scaling of the conductance $G(V)$ at the KT transition (same unit as 
in Fig. \protect\ref{Icr}). (a). For $V\gg T$, the conductance follows the 
non-equilibrium scaling $G(\protect\alpha_c,V)$. (b). For $V<T$, now the 
conductance follows the equilibrium scaling $G_{eq}(\protect\alpha_c,T)$.}
\label{GvT}
\end{figure}
\begin{eqnarray}
\label{gz-loc-app-b}
&g_{z,loc}&(\omega=0) \approx g_{z} + \frac{A^2}{2c} [1-V^{4c}]\\ \nonumber
&+& \frac{B^2}{c}V^{2c} [1-2^{-2c}],
\end{eqnarray}
 where we unambiguously identify   $A = \frac{g_{\perp}}{2} + 
\frac{cg_{\perp}}{c+|g_z|}$, $B = A V^c$; in this expression, $V$ and $\Gamma$ have been normalized to $D_0$.\\

The non-equilibrium current in the localized phase $I_{loc}(V)$ 
is obtained via the same 
approximation leading to Eq. (~\ref{Icr-app}) at the KT transition:
\begin{eqnarray}  
\label{IVapproxloc}
I_{loc}(V) &\approx & \frac{\pi V}{2} \left( \frac{\pi}{4} \left[
g_{\perp,loc}(\omega = 0)\right]^2 \right) \\ 
& +& \frac{\pi V}{2} \left( (1-\frac{\pi}{4}) \left[g_{\perp,loc}(\omega
= V/2)\right]^2 \right).  \nonumber 
\end{eqnarray}

As shown in Fig.~\ref{GvA}, we numerically 
obtain the non-equilibrium conductance 
in the localized phase.   
For very small bias voltages $V\to 0$, we find that the conductance 
reduces to the equilibrium scaling: 
$G(V)\to G_{loc}^{(e)}(T=V) \propto ({V}/{{T}^{\ast}})^{4c}$ (see 
Fig. \ref{GvA} (a) and (b)). For $g_{\perp,loc}\ll |g_{z,loc}|$ and 
$\alpha^{\ast}=\frac{1}{1+\alpha}\to 1/2$, we 
get that the exponent $4c \approx 2\alpha^{\ast} - 1$, in perfect agreement 
with that obtained in equilibrium at low temperatures: $G(T)\propto 
T^{2\alpha^{\ast} - 1}$ (Ref.~\onlinecite{zarand}). 
At higher bias voltages $0.01D_0<V< D_0
$, the conductance now follows a unique non-equilibrium form 
(consult Fig. \ref{GvB}) whose qualitative behavior is similar to that at the
KT transition. Our non-equilibrium conductance obtained numerically 
in this phase is in very good agreement with that from 
the above approximated analytical solutions in Eq.~\ref{IVapproxloc} (see Fig. \ref{GvB}). \\
\begin{figure}[!tbp]
\begin{center}
\vspace{0.5cm} 
\includegraphics[width=8.5cm]{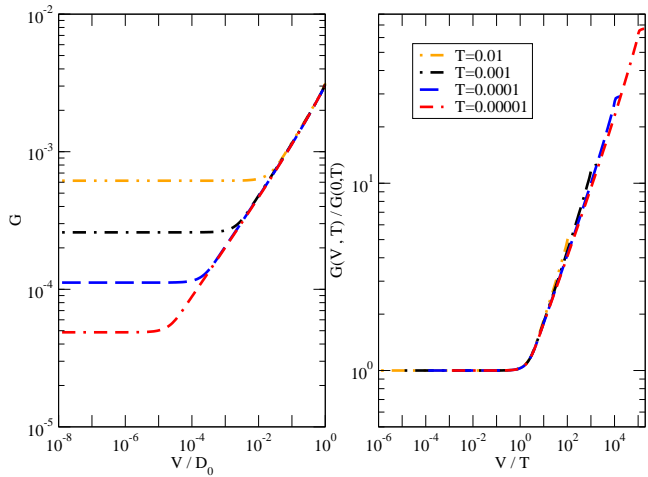}
\end{center}
\par
\vspace{-0.50cm} 
\caption{Scaling of the conductance in the localized phase with 
$g_{\perp}=0.08, g_z=-0.12$ (in unit of $D_0$). (a). For $V\gg T$, the conductance follows the 
non-equilibrium scaling $G(\protect\alpha,V)$. (b). For $V\ll T$, now the 
conductance follows the equilibrium scaling $G_{eq}(\protect\alpha_c,T)$.}
\label{GvT_loc}
\end{figure}  
\subsection{\bf Non-equilibrium conductance at finite temperatures}

We have also analyzed the finite temperature profile of the non-equilibrium 
conductance at the transition and in the localized phase. 
We distinguish two different behaviors. 
At the KT transition, for $V> T$, 
the conductance $G(V,T)$ exhibits the same non-equilibrium form as  
$T=0$, $G(V,T=0)$ (see Fig.~\ref{GvT}(a)); 
while as for $V<T$ it saturates at the value for the equilibrium 
conductance ($V=0$) at finite temperatures (see Fig.~\ref{GvT}(b). 
In the localized phase, while for $V<T$ the conductance saturates at 
$G(V=0,T)$ (Fig.~\ref{GvT_loc}(a)), 
for $V>T$, however, $G(V,T)$ exhibits universal power-law 
scaling: 
\begin{equation}
G(V,T)/ G(V=0,T) \propto (V/T)^{4 c}
\label{GVT-loc}
\end{equation} 
(see Fig.~\ref{GvT_loc}(b)). This universal power-law scaling 
behavior in $G(V,T)$ looks qualitativly similar to that from 
the recent experiment on the transport through a dissipative resonant level 
in Ref.~\onlinecite{finkelstein}. However, these two power-law 
behaviors in conductance at a finite bias and temperature 
are different in their orgins: 
The authors in Ref.~\onlinecite{finkelstein} studied the quantum critical 
behavior of a dissipative resonant level in the regime of the 
delocalized phase  
($\alpha<\alpha_c=1$). As the 
resonant level is detuned from the Fermi level, 
the system at low temperatures exhibits 
power-law scaling 
in conductance at a large bias voltage $V>T$: 
$G(V/T)\propto (\frac{V}{T})^{2\alpha}$ with $0<\alpha<1$. They showed 
further that this behavior is   
equivalent to that for a single-barrier tunneling of electrons through 
a Luttinger liquid. By contrast, the Luttinger-liquid-like 
power-law scaling in $G(V,T)$ (see Eq.~(\ref{GVT-loc}))
we find here is the generic feature of a 
dissipative resonant level in the localized phase ($\alpha>\alpha_c=1$), 
which has not yet been explored experimentally. 
Therefore, our theoretical predictions on the nonequilibrium 
transport at a dissipative 
quantum phase transition offer moltivations for further experimental 
investigations in the regime of our interest. 
The above two qualitatively different behaviors in conductance 
for $V<T$ and $V>T$ crossover at $V=T$. 
Note that similar behavior has been predicted in 
a different setup consisting of a 
magnetic Single Electron Transistor (SET) in Ref. \onlinecite{QMSi} where a 
true quantum critical point separates the Kondo screened 
and the local moment phases. 

\section{\bf Non-equilibrium finite-frequency current noise}
\begin{figure}[t]
\begin{center}
\includegraphics[width=7cm]{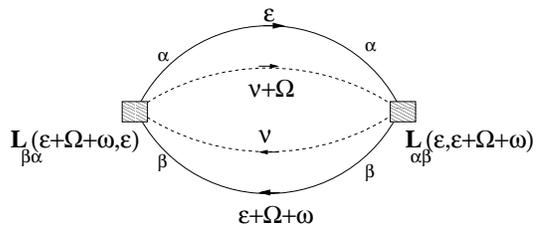}
\end{center}
\par
\vspace{-0.5cm} 
\caption{(Color online) Diagram for the FF current noise  
$S(\omega)$. The solid lines represent conduction electron propagators; 
the dashed lines denote the pseudo-fermion propagators. 
The current vertex functions $L_{\alpha\beta}(\omega_1,\omega_2)$ 
are denoted by the shaded squares.}
\label{noise_dia}
\end{figure}
In addition to non-equilibrium current and conductance 
near the localized-delocalized transition 
addressed above, further insight on the phase transition 
can be obtained  from the current fluctuations (or noise). 
 The zero frequency shot noise has been used to probe the fractional 
charge of quasiparticle excitations in FQHE  
state tunnelings\cite{glattli}. However, even 
more useful information can be found 
in the finite-frequency (FF) current 
noise, which can be used to probe the 
crossover between different quantum statistics of the 
quasiparticles\cite{statistics}. Recently, there has been theoretical 
studies on the FF current noise of a non-equilibrium Kondo 
dot\cite{schiller,schoeller,realtimeRG}. So far, 
these studies have not been extended to the non-equilibrium FF current 
noise of a dissipative quantum dot.\\
  
\subsection{Functional RG approach}

To address this issue, we combine recently developed Functional   
Renormalization Group (FRG) approach in Refs.~\onlinecite{FRG,chung2} and 
the real-time FRG approach in Ref.~\onlinecite{realtimeRG}. 
Within our FRG approaches, 
as the system moves from the delocalized to 
the localized phase, we find the smearing of the dips in  
current noise spectrum for frequencies  
$\omega\approx \pm V$; more interestingly, we find a peak-to-dip crossover 
in the AC conductance at $\omega\approx \pm V$. 
These features are detectable in experiments and 
can serve as 
alternative signatures (besides conductance) 
of the QPT in the dissipative resonant level quantum dot.\\

First, via the above mapping, 
the current through the 
dissipative resonant level quantum dot  
is given by the transverse component of the current $\hat{I}^{\perp}(t)$ 
in the effective 
anisotropic Kondo model as shown in Eq.~(\ref{currentI})\cite{chung1}. 
Following the real-time RG approach in Ref.~\onlinecite{realtimeRG} 
the Keldysh current operator through the left lead in the effective 
Kondo model (via a generalization of Eq.~(\ref{currentI})) is given by:  
$\hat{I}^{\perp}_L(t)$:
\begin{figure}[t]
\begin{center}
\includegraphics[width=7cm]{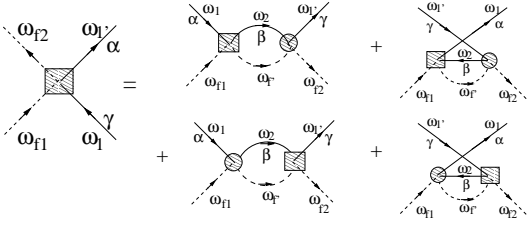}
\end{center}
\par
\vspace{-0.5cm} 
\caption{(Color online) Diagram for renormalization of the current vertex 
function $L_{\alpha\beta}(\omega_1,\omega_1^{\prime})$ (the squares). 
The solid lines represent conduction electron propagators; 
the dashed lines denote the pseudo-fermion propagators. Here, the 
Kondo couplings $g(\omega)$ are denoted by the circles. }
\label{current}
\end{figure}
\begin{eqnarray}
\hat{I}_L^{\perp}(t) &=& \frac{e}{4} \sum_{\kappa}\int dt_1 dt_2 
\sum_{\alpha,\beta}\sum dt_1 dt_2 
L_{\alpha\beta}^{\perp}(t_1-t, t-t_2)\nonumber\\ 
&\times & [s_{\alpha\beta}^{+}(t_1,t_2) S_{f}^{-}(t) + h.c.]   
\end{eqnarray}
with $\alpha,\beta=L,R$, 
$\vec{S}_f(t)=f^{\kappa\dagger}(t)\vec{\sigma}f^{\kappa}(t)$, 
$s_{\alpha\beta}^{\pm}(t_1,t_2)=c^{\kappa\dagger}_{\alpha}(t_1)\sigma^{\pm} 
c^{\kappa}_{\beta}(t_2)$. Here, 
$L_{\alpha\beta}(t_1-t, t-t_2)$ is the left  
current vertex matrix with bare (initial) matrix elements:  
$L^{0\perp}_{LL}=L^{0\perp}_{RR}=0$, $L_{LR}^{0\perp}=-L_{RL}^{0\perp}= 
{\it i} g_{LR}^0 \equiv g_{\perp}$, $L^{0z}_{LL}=L^{0z}_{RR}\equiv g_z$, 
$L_{LR}^{0z}=-L_{RL}^{0z}= 0$, and $\kappa=\pm 1$ being the upper and lower 
Keldysh contour, respectively. The emission component of the 
non-equilibrium FF noise of a Kondo quantum dot, $S^<(t)$, is given by 
the current-current correlator:
\begin{equation}
S^<_{LL}(t) \equiv \langle\hat{I}^{\perp}_L(0) \hat{I}^{\perp}_L(t)\rangle
\end{equation} 

Similarly, the absorption part of the noise is defined as: 
$S^>(t)\equiv \langle\hat{I}^{\perp}_L(t) \hat{I}^{\perp}_L(0)\rangle$. 
Note that the current operator $\hat{I}^{\perp}_L(t)$ is non-local in 
time under RG; the current vertex function $L_{\alpha\beta}(t_1-t, t-t_2)$ 
therefore acquires the double-time structure: it keeps track of not only 
the times electrons enter ($t_1$) and leave ($t_2$) the dot, 
but also the time $t$ at which the current is measured\cite{realtimeRG}. 
The double-time structure of the current operator 
automatically satisfies the current conservation: 
$\hat{I}^{\perp}_L(t) = - \hat{I}^{\perp}_R(t)$ (Ref.~\onlinecite{realtimeRG}). 
\begin{figure}[t]
\begin{center}
\includegraphics[width=8.5cm]{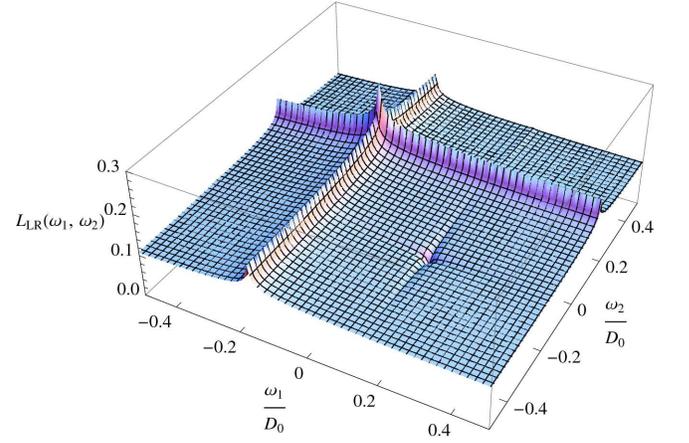}
\end{center}
\par
\vspace{-0.5cm} 
\caption{(Color online) 3D plot for $L_{LR}(\omega_1, \omega_2)$ 
at zero temperature  
in the delocalized phase with bare Kondo couplings being 
$g_{\perp}^{0} =0.05 D_0$, $g_{z}^{0}=0.05 D_0$. 
The bias voltage is fixed at $V=0.32 D_0$.}
\label{L12_deloc}
\end{figure}
The frequency-dependent 
current noise $S(\omega)$  
is computed via the second-order renormalized 
perturbation theory (see diagram in Fig.~\ref{noise_dia}). 
Note that due to the double-time structure of the current vertex function 
$L_{\alpha\beta}(t_1,t_2)$, in the Fourier (frequency) space,  
$L_{\alpha,\beta}(\epsilon+\omega,\epsilon)$ 
has a two-frequency structure; it 
depends on the incoming ($\epsilon+\omega$) 
and outgoing ($\epsilon$) frequencies of the electron 
(see Fig.~\ref{noise_dia}). The result reads:
\begin{eqnarray}
S^{<}(\omega) &=& \sum_{\alpha,\beta=L,R} -2 
Re(D_{\alpha\beta}(\omega)^{<})
\end{eqnarray} 
where the correlator 
$D_{\alpha\beta}(\omega)$ is computed by the diagram in Fig.~\ref{noise_dia}:
\begin{eqnarray}
D_{\alpha\beta}(\omega)^{<} &=& \int \frac{d\Omega}{2\pi}
[\chi_{\alpha\beta}(\Omega,\omega) \chi_f(\Omega)]^<,\nonumber \\ 
\chi_{\alpha\beta}(\Omega,\omega) &=& \int \frac{d\epsilon}{2\pi} 
\hat{G}_{\alpha}(\epsilon)\hat{G}_{\beta}(\epsilon+\Omega+\omega)\nonumber \\
&\times& L_{\alpha\beta}^{\perp}(\epsilon+\omega, \epsilon)  
L_{\beta\alpha}^{\perp}(\epsilon, \epsilon+\omega),\nonumber \\
\chi_f(\Omega) &=& \int \frac{d\nu}{2\pi} \hat{G}_f(\nu) \hat{G}_f(\nu+\Omega), 
\end{eqnarray}
where $\hat{G}$ is the Green's function in $2\times 2$ Keldysh space, and its 
lesser and greater Green's function 
are related to its retarded, advanced, and Keldysh components by:
\begin{eqnarray}
G^< &=& (G^K - G^R + G^A)/2\nonumber \\
G^> &=& (G^K +G^R - G^A)/2
\end{eqnarray}
The lesser ($G^<$) and greater ($G^>$) components of Green's function 
of the conduction electron in the leads and of the quantum dot (impurity) 
are given by:
\begin{eqnarray}
G_{L/R}^<(\epsilon) &=& {\it i} A_c(\epsilon) f_{\epsilon-\mu_{L/R}}\nonumber \\
G_{L/R}^>(\epsilon) &=& {\it i} A_c(\epsilon) (1-f_{\epsilon-\mu_{L/R}})\nonumber \\
G_{f\sigma}^<(\epsilon) &=& 2\pi {\it i} \delta(\epsilon) n_{f\sigma}(\epsilon)\nonumber \\
G_{f\sigma}^>(\epsilon) &=& 2\pi {\it i} \delta(\epsilon) (n_{f\sigma}(\epsilon) -1), 
\end{eqnarray}
where $A_c(\epsilon)= 2\pi N_0 \Theta(D_0- \epsilon)$ 
is the density of states of the leads, 
$n_{f\sigma}(\epsilon)=\langle f^{\dagger}_{\sigma}f_{\sigma}\rangle$ is 
the occupation number of the pseudofermion which 
obeys $n_{f\uparrow}+n_{f\downarrow}=1$, 
$n_{f\sigma}(\epsilon\rightarrow 0)= 1/2$ in the delocalized phase and 
$n_{f\uparrow}(\epsilon\rightarrow 0)\rightarrow 0$, 
$n_{f\downarrow}(\epsilon\rightarrow 0)\rightarrow 1$ in the localized 
phase\cite{chung1,latha}. Here, the pseudofermion occupation number 
$n_{f\sigma}$ and the occupation number on the dot 
$n_d$ are related via $\langle n_{f\uparrow}-n_{f\downarrow}\rangle= \langle 
n_d\rangle -1/2$ 
(Refs.~\onlinecite{chung1,chung2}). The renormalized current vertex 
function $L_{\alpha\beta}^{\perp}(\omega_1, \omega_2)$ and the Kondo couplings 
$g_{\perp}(\omega)$, $g_z(\omega)$ are obtained from the 
non-equilibrium Functional RG approaches in Ref.~\onlinecite{realtimeRG} and 
Refs.~\onlinecite{FRG,noneqRG}, respectively. 
Carrying out the calculations, the finite-frequency noise spectrum reads:
\begin{figure}[t]
\begin{center}
\includegraphics[width=8.5cm]{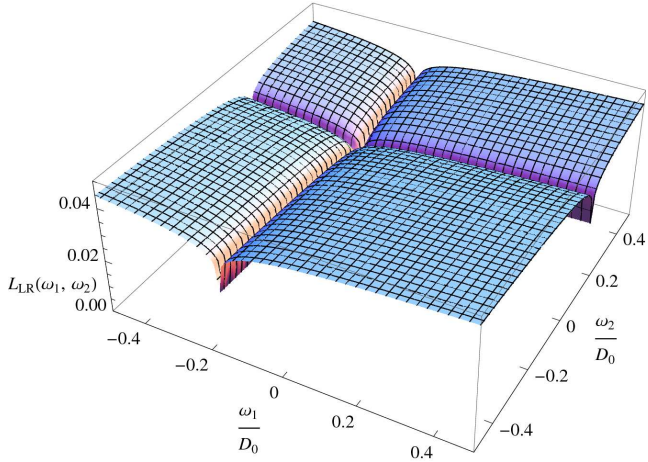}
\end{center}
\par
\vspace{-0.5cm} 
\caption{(Color online) 3D plot for $L_{LR}(\omega_1, \omega_2)$ 
at zero temperature in the localized phase with bare Kondo couplings being 
$g_{\perp}^{0} =0.05 D_0$, $g_{z}^{0}=-0.1 D_0$. 
The bias voltage is fixed at $V=0.32 D_0$.}
\label{L12_loc}
\end{figure}
\begin{eqnarray}
S^{<}(\omega) &=& \sum_{\alpha,\beta=L,R}\frac{3}{8}\int d\epsilon 
L_{\alpha\beta}^{\perp}(\epsilon+\omega, \epsilon)  
L_{\beta\alpha}^{\perp}(\epsilon, \epsilon+\omega)\nonumber \\
&\times& f_{\epsilon-\mu_\alpha}(1-f_{\epsilon-\mu_\beta}), 
\label{Snu}
\end{eqnarray}
where $f_{\epsilon-\mu_\alpha}$ is the Fermi function of the lead $\alpha=L/R$ 
given by $f_{\epsilon-\mu_\alpha} = 1/(1+e^{(\epsilon-\mu_{\alpha})/k_BT})$. 
The symmetrized noise spectrum reads:
\begin{equation}
S(\omega) = \frac{1}{2}[S^{<}(\omega) + S^{>}(\omega)]
\end{equation} 
with the relation between emission and absorption parts of the 
noise spectrum in frequency space $S^<(\omega) = S^>(-\omega)$ being used.

The frequency-dependent Kondo couplings $g_{\perp,z}(\omega)$
and current vertex functions $L^{\perp}_{\alpha\beta}(\omega_1,\omega_2)$ 
are obtained self-consistently within the FRG approaches, 
which can be divided into two parts. 
First, the Kondo couplings $g_{\perp,z}(\omega)$ are solved via 
Eq.~(\ref{gpergz})\cite{noneqRG,FRG,chung2} 
together with the generalized frequency-dependent 
dynamical decoherence rate $\Gamma(\omega)$ appearing in 
$\Theta _{\omega }=\Theta (D-|\omega +\mathit{i}\Gamma(\omega) |)$ in 
Eq.~(\ref{gpergz}). Here, $\Gamma(\omega)$ is 
obtained from the imaginary part of the pseudofermion self 
energy\cite{decoherence,FRG,chung2}:
\begin{eqnarray}
\Gamma(\omega) =  
\frac{\pi}{4} &\int{d\epsilon}& g_{\perp}(\epsilon+\omega) 
g_{\perp}(\epsilon) 
[f^{L}_{\epsilon}-f^{R}_{\epsilon+\omega}]\nonumber \\
&+ & g_{z}(\epsilon+\omega) 
g_{z}(\epsilon) 
[f^{L}_{\epsilon}-f^{L}_{\epsilon+\omega}]\nonumber \\
&+& (L\rightarrow R).
\label{gamma}
\end{eqnarray}
Note that the zero-frequency decoherence rate 
$\Gamma(\omega=0)$ corresponds to the decoherence rate 
$\Gamma$ obtained in Eq.~(\ref{gammaV})\cite{noneqRG}. 
We have solved the RG equations Eq.~(\ref{gpergz}) 
subject to Eq. ~(\ref{gamma}) self-consistently\cite{footnote}.\\

The solutions for 
$g_{\perp}(\omega)$, $g_{z}(\omega)$ and $\Gamma(\omega)$ close to 
the KT transition 
are shown in Refs.~\onlinecite{chung1,chung2} 
(see also Fig.~\ref{gamma_omega} and Fig. ~\ref{gpergzfig}). 
As the system 
goes from the delocalized to localized phase, 
the features in $g_{\perp}(\omega)$ at $\omega=\pm V/2$ 
undergoes a crossover from symmetric two peaks to symmetric two dips, 
while the symmetric two peaks in $g_{z}(\omega=\pm V/2)$ still 
remain peaks. 
 The finite-frequency non-equilibrium 
decoherence rate $\Gamma(\omega)$ monotonically increases with 
increasing $\omega$, it shows  logarithmic sigularities 
 at $|\omega|=V$ in the delocalized phase\cite{chung2}. 
As the system moves to the localized phase, the overall 
magnitude of $\Gamma(\omega)$ decreases rapidly and the 
singular behaviors at $\omega=\pm V$ get smeared out\cite{chung2}.\\
\begin{figure}[t]
\begin{center}
\includegraphics[width=7.5cm]{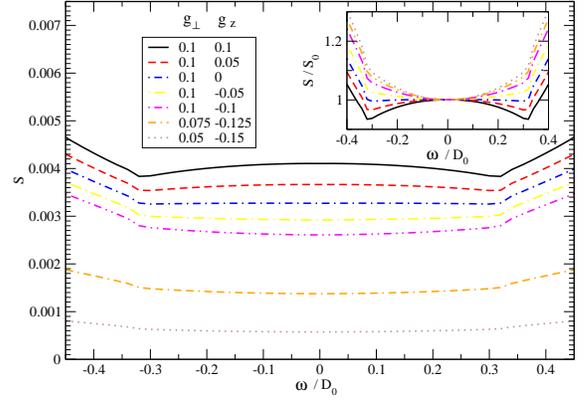}
\end{center}
\par
\vspace{-0.5cm} 
\caption{(Color online) $S(\protect\omega)$ 
at zero temperature versus $\omega$ across the KT transition. 
The bias voltage 
is fixed at $V=0.32 D_0$. Inset:  $S(\protect\omega)$  
at zero temperature versus $\omega$ normalized to $S_0 = S(\omega=0)$.}
\label{S_scaled}
\end{figure}
Next, following Ref.~\onlinecite{realtimeRG}, 
we generalize the RG scaling equation for the general current 
vertex function $L_{\alpha\beta}(\omega)$ for the 
anisotropic Kondo model (see diagrams in Fig.~\ref{current} and also in 
Fig. 1 of Ref.~\onlinecite{realtimeRG}). 
The RG scaling equations for 
the general vertex functions $L_{\alpha\beta}^{\perp,z}(\omega_1,\omega_2)$ 
can be simplified as:
\begin{eqnarray}
\frac{d L_{\alpha\beta}(\omega_1,\omega_2)}{d ln D} &=& \sum_{\gamma=L,R} 
L_{\alpha\gamma}(\omega_1,\omega_2)\Theta_{\mu_\gamma}(\omega_2) 
g_{\gamma\beta}(\omega_2) \nonumber\\ 
&+& g_{\alpha\gamma}(\omega_1)\Theta_{\mu_\gamma}(\omega_1)
L_{\gamma\beta}(\omega_1,\omega_2)
\label{Lij}
\end{eqnarray}
where we make the following 
identifications: 
$g_{LR/RL}(\omega) \rightarrow g_{LR/RL}^{\perp}(\omega) 
\equiv g_{\perp}(\omega)$,  
$g_{\alpha\alpha}(\omega)\rightarrow g_{LL/RR}^z(\omega) 
\equiv g_z(\omega)$. Similarly,  
$L_{LR/RL}(\omega_1,\omega_2) \rightarrow L_{LR/RL}^{\perp}(\omega_1,\omega_2)$ 
refers to only the transverse component of the current vertex function 
$L_{\alpha\beta}(\omega_1,\omega_2)$; while   
$L_{LL/RR}\rightarrow L_{LL/RR}^z$ refers only to the longitudinal 
part of $L_{\alpha\alpha}$. 
Here, the frequency-dependent Kondo couplings $g_{\perp, z\sigma}(\omega)$ in 
Eq.~(\ref{Lij}) are obtained from Eq.~(\ref{gpergz}) and Eq.~(\ref{gamma}). 
Note that the scaling equations for $L_{\alpha\beta}(\omega_1,\omega_2)$ 
via Ref.~\onlinecite{realtimeRG} 
can also be expressed within the RG 
approach in Ref.~\onlinecite{noneqRG} via a straightforward generalization by 
allowing for the two-frequency dependent 
vertex functions $L_{\alpha\beta}(\omega_1,\omega_2)$ 
where $\omega_{1(2)}$ refers to the incoming (outgoing) frequency 
(see Fig.~\ref{noise_dia} and Fig.~\ref{current}). \\ 
\begin{figure}[t]
\begin{center}
\includegraphics[width=8.50cm]{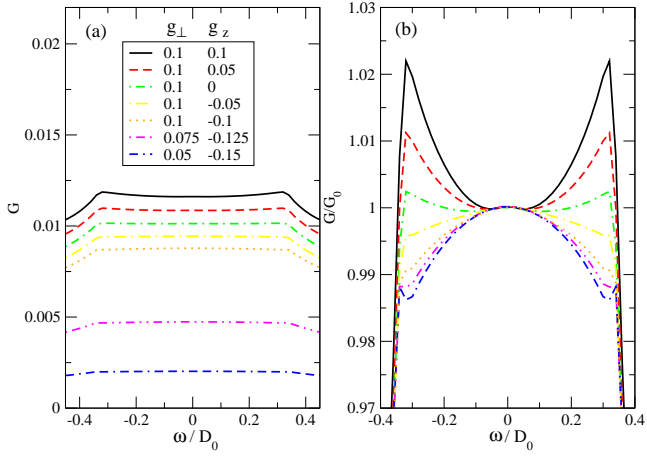}
\end{center}
\par
\caption{(Color online) (a) The zero-temperature AC conductance 
$G_{AC}(\protect\omega)$ defined in Eq. (~\ref{GACeq}) 
versus $\omega$ across the KT transition. (b) $G_{AC}(\protect\omega)$ 
at zero temperature  
versus $\omega$ normalized to $G_0 \equiv G_{AC}(\omega=0)$.  
The bias voltage is fixed at $V=0.32 D_0$.}
\label{GAC}
\end{figure}

\subsection{Results}

We solved the self-consistent 
RG scaling equations Eq.~(\ref{Lij})  
for the current vertex functions 
with the help of the solutions for the renormalized Kondo couplings 
via Eq.~(\ref{gpergz}) and Eq.~(\ref{gamma}). 
The typical results at zero temperature 
are shown in Fig.~\ref{L12_deloc} and Fig.~\ref{L12_loc}; 
they exhibit 
the following symmetry:  
$L_{\alpha\beta}(\omega_1,\omega_2) = -L_{\beta\alpha}(\omega_2,\omega_1)$. 
Note that since the initial conditions for the current vertex function 
have the following structures: $L_{\alpha\alpha}^0=0$, $L_{LR}^0\neq 0$, 
we find $L_{\alpha\alpha}(\omega_1,\omega_2)\ll L_{LR}(\omega_1,\omega_2)$. 
In the delocalized (Kondo) phase, a sharp peak is developed in 
$L_{LR}(\omega_1,\omega_2)$ for 
$(\omega_1,\omega_2)=(V/2,-V/2)$; while as a small dip is formed 
for $(\omega_1,\omega_2)=(-V/2,V/2)$. Meanwhile, in general 
$L_{LR}(\omega_1,\omega_2)$ is maximized at $\omega_{1(2)}=\pm V/2$ 
for fixed $\omega_{2(1)}$. This agrees perfectly with the result 
in Ref.~\onlinecite{realtimeRG}. In the localized phase, however, 
we find the opposite: 
$L_{LR}(\omega_1,\omega_2)$ develops a sharp dip at 
$(\omega_1,\omega_2)=(V/2,-V/2)$; and it is minimized 
$\omega_{1(2)}=\pm V/2$ for fixed $\omega_{2(1)}$. The peak-dip 
structure of the current vertex function $L_{\alpha\beta}$ 
plays a crucial role in determining the noise spectrum 
both in delocalized and in the localized phases.\\ 

Substituting the numerical solutions for $L_{\alpha\beta}(\omega_1,\omega_2)$ 
and $g_{\alpha\beta}(\omega)$ into Eq.~(\ref{Snu}), 
we get the zero-temperature FF noise $S(\omega)$. The results at 
zero temperature are shown 
in Fig.~\ref{S_scaled}. 
First, the overall magnitude of $S(\omega)$ 
decreases rapidly as the system crosses over from the delocalized 
to the localized phase. This can be understood easily as the current 
decreases rapidly in the crossover, leading to a rapid decrease in 
the magnitude of noise.   
For $|\omega| > V$, $S(\omega)$ in both phases 
increases monotonically with increasing $\omega$ 
due to the increase of the photon emission at higher 
energies\cite{realtimeRG}. 
For $|\omega| \le V$, however, it 
changes from a 
peak to a dip centered at $\omega =0$ as 
the system crosses overs from delocalized to localized phase 
(see Fig.~\ref{S_scaled}). 
At $|\omega|=V$, $S(\omega)$ exhibits a dip (minima) in the delocalized phase, 
a signature of the non-equilibrium Kondo effect; 
while as the system crosses over to the localized phase 
the dips are gradually  smeared out and they change into  
a ``kink''-like singular point at $\omega = \pm V$, connecting 
two curves between $\omega<V$ and $\omega>V$.\\ 
\begin{figure}[t]
\begin{center}
\vspace{0.6cm}
\includegraphics[width=9cm]{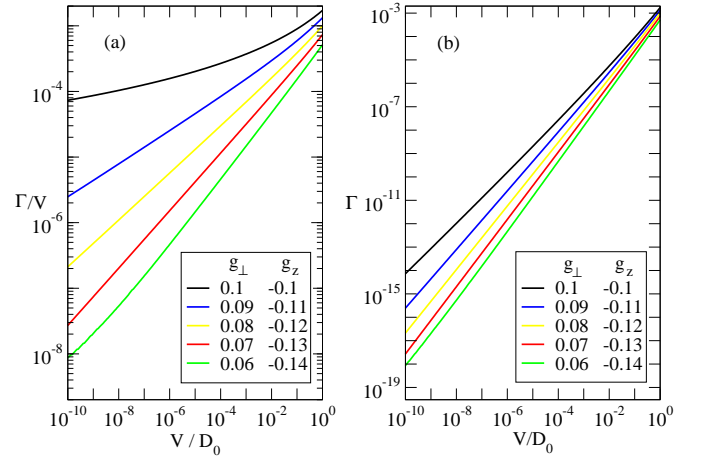}
\end{center}
\par
\vspace{-0.3cm} 
\caption {(a) $\frac{\Gamma}{V}$ and (b) $\Gamma$ as a function of $V/D_0$ near the KT transition. }
\label{gammaVfig}
\end{figure}
We furthermore computed the non-equilibrium AC conductance at 
zero temperature\cite{safi,realtimeRG}: 
\begin{equation}
G_{AC}(\omega)= \frac{S^{<}(\omega)-S^{>}(\omega)}{\omega}
\label{GACeq}
\end{equation} 
across the transition. 
Note that $G(\omega=0)=dI/dV$ corresponds to the non-equilibrium 
differential conductance. As shown in Fig.~\ref{GAC} (a), 
in the delocalized phase 
the splitted peaks in $G_{AC}(\omega)$ at $\omega=\pm V$ are 
signatures of the Kondo resonant at finite bias, and are 
consistent with the dips at seen in the noise spectrum. As the system 
moves to the localized phase, the overall magnitudes of $G_{AC}(\omega)$ 
as well as the pronounced splitted 
Kondo peaks at $\omega=\pm V$ get suppressed; they change into  
dips deep in the localized phase (see Fig.~\ref{GAC} (b)). 
In response to 
this change in the splitted Kondo peaks, the overall shape 
of $G_{AC}(\omega\rightarrow 0)$ shows a dip-to-hump crossover 
near $\omega=0$. Note that the suppression of the Kondo 
peaks for $G_{AC}(\omega)$ 
at $\omega=\pm V$ corresponds to the smearing of the 
dips at $\omega=\pm V$ shown in the noise spectrum $S(\omega)$ 
(see Fig.~\ref{S_scaled}). 
The above 
evolution in the noise spectrum matches well with the non-equilibrium transport 
properties studied in Refs.~\onlinecite{chung1,latha}, and can serve 
as alternative signatures  
of the localized-delocalized transition in future experiments. \\

\section{\bf Discussions}

We would like to make a few remarks before we conclude. 
Firstly, the distinct non-equilibrium scaling behavior seen here is 
in fact closely 
tied to the non-trivial (non-linear) $V$ dependence of the decoherence rate 
$\Gamma (V)$ which cuts off the RG flow 
(see Fig. \ref{gammaVfig} (a) and (b)). 
The decoherence rate $\Gamma$ near the transition 
clearly plays a very different role as compared to  
the temperature near the transition. 
In particular, at $T=0$ we find that $\Gamma \sim 
\frac{1}{2}I$ is a highly non-linear function in $V$, resulting in the 
observed deviation of the non-equilibrium scaling from that in equilibrium. 
In fact, we can obtain the analytical form via the approximation in 
Eq.~(\ref{IVapprox}) and Eq.~(\ref{IVapproxloc}). 
At the KT transition, $\Gamma/V$ 
shows a logarithimic decrease as $V$ decreases 
(see Eqs.~(\ref{Icr-app}), (\ref{gcr-app})); 
while in the localized phase 
it exhibits a combined power-law and logarithmic dependence 
on $V$ (see Eqs.~(\ref{gp-loc-app-a}), (\ref{gp-loc-app-b}), 
(\ref{gz-loc-app-a}), 
(\ref{gz-loc-app-b}), (\ref{IVapproxloc})). \\

By contrast, the equilibrium decoherence rate $\Gamma(V=0,T)$ 
shows a clear power-law behavior in the localized phase  
at low temperatures, $T\rightarrow 0$ (see Eq.~(\ref{GT-loc})):
\begin{equation}
\Gamma(V=0,T) \propto \left(\frac{T}{T^{\ast}}\right)^{1+4c}, 
\end{equation}
which is consistent with the prediction made in Ref.~\onlinecite{hur-lifetime-lutt} 
for the electron lifetime in Luttinger liquids.

Meanwhile, at the KT 
transition and in the localized phase, since $\Gamma\ll V$, 
the RG flow for $g_{\perp/z}$ are cut off at an energy scale $\Gamma$ 
much lower than $V$, leading to smaller renormalized 
couplings $g_{\perp/z}$ in magnitude 
compared to their corresponding equilibrium values 
$g^{(e)}_{\perp/z}(T=V)$, $|g_{\perp/z}|< |g^{(e)}_{\perp/z}(T=V)|$.  
This results in smaller conductance 
than that in equilibrium, $G(V)<G^{eq}(T)$.\\

Secondly, it is of fundamental importance to study further 
the possible scaling behaviors 
in non-equilibrium dynamical quantities near the transition, 
such as the $\omega/T$ scaling in dynamical charge susceptibility at the KT 
transition and in the localized phase. In particular, the question has been 
raised on the existence of 
the concept of ``effective temperature'' that allows one to extend 
the fluctuation-dissipation theorem in equilibrium 
to the non-equilibrium (non-linear) regime\cite{Stefan}. 
It is also interesting to address the 
crossover between delocalized phase with $G(V)\propto 1/\ln^2(V/T_K)$ 
where $\ln T_{K}\propto 1/(\alpha-\alpha_c)$ to KT point with 
$G(V) \propto 1/\ln^2(T/\mathcal{D})$ and further 
to the localized phase with power-law conductance $G(V)\propto V^{\beta}$. 
To date, the full crossover function of the conductance is not known yet. 
Further study is therefore needed to investigate these issues.

\section{\bf Conclusions}

In summary, we have investigated the non-equilibrium transport at a 
QPT using a standard nano-model, the dissipative resonant level model. 
By employing an exact mapping onto the anisotropic Kondo model and by 
applying a controlled energy-dependent RG and Functional RG approaches 
to our model system we have 
calculated the renormalized coupling functions $g_{\perp,z}(\omega)$, 
the decoherence rate $\Gamma$, the current $I$, differential conductance 
$G(V,T)$, and the current noise spectrum $S(\omega)$. 
For $V\rightarrow 0$, the conductance $G$ follows the equilibrium behavior; 
by increasing $V$, the 
frequency-dependence of the couplings begins to play an important role and 
therefore we systematically find scaling behavior of the non-equilibrium 
conductance very distinct from that of the equilibrium counterpart. 
We have also 
analyzed the finite temperature profile of $G(V,T)$ at the transition as 
well as in the localized phase and  
found that the conductance shows different behaviors for $V>T$ and $V<T$; it 
exhibits $V/T$ scaling behavior for $V\ll T$.\\ 

Regarding transport properties of our system near the 
transition, the role played by the 
bias voltage is very different from that played 
by the temperature. The key to these very different behaviors lies in the 
fact that the non-equilibrium charge (or effective spin) decoherence rate, 
which serves as a cutoff for the RG flows of the Kondo couplings, is a 
highly non-linear function of the bias voltage. 
Further investigations are needed to address the full crossover function 
in conductance as well as the scaling behaviors 
of the dynamical quantities near the transition in a search for the existence 
of the ``effective temperature'' that allows one to generalize the equilibrium 
fluctuation-dissipation theorem to the non-equilibrium regime. 
Furthermore, we provide signatures of the localized-delocalized transition 
in the finite-frequency current noise spectrum and the AC conductance.  
Our results have a direct experimental relevance for dissipative two-level 
systems; moreover, they are applicable for describing non-equilibrium 
transport of a resonant level coupled to interacting chiral Luttinger liquid 
generated by fractional quantum Hall edge states via the mappings discussed 
in Appendix A. Finally, our model system has direct relevance for the recent 
experiment in a quantum dot coupled 
to resistive environment as shown in Ref.~\onlinecite{finkelstein}. Our work 
motivates future experimental 
as well as theoretical investigations on 
dissipative quantum phase transitions in nanosystems.\\

\acknowledgments 

We are grateful to D. Goldhaber-Gordon, P. Moca, G. Zarand,  
and S. Kirchner for stimulating discussions, and to 
R.T. Chang and K.V.P. Latha for technical support. 
We acknowledge the generous support from the NSC grant 
No.98-2112-M-009-010-MY3, No.101-2628-M-009-001-MY3, 
the MOE-ATU program, the CTS of NCTU, the NCTS of  
Taiwan, R.O.C. (C.H.C.), 
the Department of Energy in USA under the contract DE-FG02-08ER46541 (K.L.H.), 
the German-Israeli Foundation via G 1035-36.14/2009, 
the DFG via FOR 960 (M.V.,P.W.), GRK 1621 (M.V.), and the Center for 
Functional Nanostructures CFN (P.W.), the U.S. Department of Energy,
Office of Basic Energy Sciences, Division of Materials Sciences
and Engineering under Award DE-SC0002765 (G.F.). 
C.H.C. has benefitted from the 
visiting programs of KITP, ICTP, and MPI-PKS. K.L.H. acknowledges KITP for 
hospitality. G.F. thanks NCTS, NCTU of Taiwan R.O.C. for hospitality during 
the visit. 


\appendix

\section{Useful mappings}

In this Appendix, we provide detailed derivations 
on various mappings mentioned in Sec. II. Via bosonization 
and refermionization techniques,  
the three mappings described below will follow one from the other, 
but there are a few technical details that will change.

\subsection{Mapping a dissipative resonant level model onto anisotropic Kondo model}

We describe in details on the mapping of 
dissipative resonant level model in Eq.~(\ref{RLM}) onto 
anisotropic Kondo model 
in Eq.~(\ref{Hkondo}). Our goal is to connect the parameters of 
these two equations in the main text.

We first start from Eq.~(\ref{RLM}):
\begin{eqnarray}
H &=& \sum_{k,i=1,2} (\epsilon(k)-\mu_i) c^{\dagger}_{k i}c_{k i} + t_{i}
c^{\dagger}_{ki} d + h.c. \\
&+& \sum_{r} \lambda_{r} (d^{\dagger}d-1/2) (b_{r} + b^{\dagger}_{r}) +
\sum_{r} \omega_{r} b^{\dagger}_{r} b_{r},  \nonumber
\label{RLM-appendix}
\end{eqnarray}
where $t_{i}$ is the (real-valued) hopping amplitude between the 
lead $i$ and the quantum dot, $c_{ki}$ and $d$ are electron operators for 
the (Fermi-liquid type) leads and the quantum dot, respectively. 
$\mu_i = \pm V/2$ is the chemical potential applied on the lead $i$ 
($V$ denotes the bias voltage), while the dot level is at zero 
chemical potential. Here, $b_{r}$ are the boson operators of the 
dissipative bath with an Ohmic type spectral density. 
It proves to be more convenient to re-express 
the dissipative boson fields $b_r$ and $b^{\dagger}_r$ in terms of 
the canonical fields $\hat{\phi}_0(x,t)$ 
and $\hat{\Pi}_0(x,t)$ as: \cite{delft,lehur1}:
\begin{eqnarray}
\hat{\phi}_0(x,t) &=& \int_{-\infty}^{\infty} \frac{dp}{2\pi \sqrt{2|p|}} 
[b_p e^{{\it i} p x} + b^{\dagger}_p e^{{\it i}p x}] e^{-a |p|/2}\nonumber \\
\hat{\Pi}_0(x,t) &=& \partial_t \hat{\phi}_0(x,t)\nonumber \\
\end{eqnarray}
where $\omega_r = v_b p_r$ with $v_b$ being the phonon velocity, and  
the boson fields $\hat{\phi}_0(x,t)$ and 
$\hat{\Pi}_0(x,t)$ satisfies the commutation relation: 
$[\hat{\phi}_0(x,t), \hat{\Pi}_0(x^{\prime},t)] = 
{\it i} \delta (x-x^{\prime})$. 
The dissipative boson bath can therefore be re-expressed as: 
\begin{eqnarray}
H_{diss} &=& \sum_{r} \omega_{r} b^{\dagger}_{r} b_{r} 
= \int \frac{dp}{2\pi} |p| b^{\dagger}_p b_p \nonumber \\
&=& \frac{1}{2} \int dx [(\partial_x \hat{\phi}_0)^2(x,t) + \hat{\Pi}_0^2(x,t)].
\label{RLM-diss}
\end{eqnarray}
Here, the velocity of the boson field $\hat{\phi}_0$ is set to be $1$. 

We start the mapping by bosonizing the fermionic operators in the leads: 
\begin{equation}
c_{\alpha}(0) = \frac{1}{\sqrt{2\pi a}} F_{\alpha}  e^{{\it i}\varphi_{\alpha}(0)}, 
\end{equation}
where we have introduced the (standard) Klein factors $F_{\alpha}$ ensuring anti-commutation relations and $a$ is a short-distance cutoff (lattice spacing). 
The fermionic baths of conduction electrons can be re-written as:
\begin{eqnarray}
H_{leads} &=&\sum_{k,i=1,2} (\epsilon(k)-\mu_i) c^{\dagger}_{k i}c_{k i}\nonumber \\
&=&  \frac{1}{2} \int dx \sum_{\alpha=1,2}
[(\partial_x \varphi_\alpha)^2(x,t) + \Pi_\alpha^2(x,t)]
\label{RLM-leads}
\end{eqnarray}
where the Fermi velocity of the electrons is set to be $1$.

The level on the quantum dot can be mapped onto a pseudo-spin: $d = F_d S^{-}$ 
and $S^z= d^{\dagger}d-1/2$; 
$\alpha=1,2$ represent the two leads. The coupling between the dot and the 
dissipation bath ($\lambda_i$ term) can be absorbed in the 
tunneling part of the Hamiltonian through the unitary transformation $U_B$\cite{lehur1}: 
\begin{eqnarray}
{U}_B &=& e^{{\it i}\sqrt{\frac{1}{K_b}} S_z \hat{\phi}_0} \\ \nonumber
\tilde{H_t} &=& U_B^{\dagger} H_t U_B \\ \nonumber
&=& \sum_{\alpha=1,2} t_{\alpha}  F_{\alpha}^{\dagger} F_d
e^{{\it i} \sqrt{\frac{1}{K_b}} \hat{\phi}_0 } e^{{\it i} \varphi_{\alpha}(0)} S^{-} + H.c. 
\label{Ub}
\end{eqnarray}
with $K_b\equiv \frac{1}{\alpha}$. Here, $\alpha$ refers to the strength 
of the coupling between the resonant level and the dissipative boson bath, 
and we set $2\pi a = 1$ for simplicity.

We can simplify our variables even further by combining the above fields describing the leads and the noise: 
$\tilde{\phi}_{s,\alpha} = \sqrt{K} (\varphi_{\alpha} + 
\sqrt{\frac{1}{K_b}}\hat{\phi}_0)$,
$\tilde{\phi}_{a,\alpha} = \sqrt{K} (\sqrt{\frac{1}{K_b}}\varphi_{\alpha} -  
\hat{\phi}_0)$, where  $\frac{1}{K} = \frac{1}{K_b} + 1
= \alpha +1 \equiv \frac{1}{\alpha^{\ast}}$. Note that here 
$K$ may be interpreted 
as the effective Luttinger liquid parameter as the effect of 
Ohmic dissipation 
on the quantum dot plays a similar role as interactions in the 
Luttinger liquid leads coupled to the dot with the identification $K = \frac{1}{1+\alpha}$. The comined bosonic and fermionic bath $\tilde{H}_{bath}$ 
can be re-expressed in terms of these new boson fields:
\begin{eqnarray}
\tilde{H}_{bath} &\equiv& H_{leads} +H_{diss}\nonumber \\
&=& \frac{1}{2} \int dx \sum_{\alpha=1,2}
[(\partial_x \varphi_\alpha)^2(x,t) + \Pi_\alpha^2(x,t)] \nonumber \\
&+& \frac{1}{2} \int dx [(\partial_x \hat{\phi}_0)^2(x,t) 
+ \hat{\Pi}_0^2(x,t)]\nonumber \\
&=& \frac{1}{2} \int dx \sum_{\alpha=1,2}[(\partial_x \tilde{\phi}_{s,\alpha})^2(x,t) 
+ \hat{\Pi}_{s,\alpha}^2(x,t)\nonumber \\
&+& (\partial_x \tilde{\phi}_{a,\alpha})^2(x,t) 
+ \hat{\Pi}_{a,\alpha}^2(x,t)],
\label{tilde-H-bath}
\end{eqnarray}
where $\Pi_{s(a),\alpha}$ fields are canonically conjuate to 
the fields $\tilde{\phi}_{s(a),\alpha}$. Note that as we shall see below 
only the fields from symmetric combinations  
$\tilde{\phi}_{s,\alpha}$ and $\Pi_{s,\alpha}$ couple to 
the tunneling and chemical potential terms, the antisymmetric 
combinations $\tilde{\phi}_{a,\alpha}$ and $\Pi_{a,\alpha}$ are de-coupled 
from the rest of the Hamiltonian.  

The tunneling and chemical potential parts of the Hamiltonian now become: 
\begin{eqnarray}
\tilde{H_t} &=& U_B^{\dagger} H_t U_B = \sum_{\alpha=1,2} t_{\alpha}F_{\alpha}^{\dagger} F_d
e^{{\it i} \frac{\tilde{\phi}_{s,\alpha}}{\sqrt{K}} } S^{-} + H.c. 
 \\ \nonumber
 \tilde{H}_{\mu} &=& U_B^{\dagger} H_{\mu} U_B = -\frac{V}{2} \sqrt{\frac{1}{K}} 
(\partial_x \tilde{\phi}_{s,1} - \partial_x \tilde{\phi}_{s,2} )
\label{KondobeforeU1U2}
\end{eqnarray}

Close to $\alpha^*=1/2$ (transition), we can map our model onto the 2-channel aniotropic Kondo model. 
After applying the two unitary transformations ${U}_1 = e^{{\it i} (\frac{\tilde{\phi}_{s,1}}{\sqrt{K}} 
- \sqrt{2}\tilde{\phi}_{s,1} )S_z }$ and ${U}_2 = e^{{\it i} (\frac{\tilde{\phi}_{s,2}}{\sqrt{K}} 
- \sqrt{2}\tilde{\phi}_{s,2} ) S_z }$, we obtain:
\begin{eqnarray}
\tilde{H}_t^{''} &=& U_2^{\dagger} U_1^{\dagger} \tilde{H}_t U_1 U_2 \\ \nonumber
&=&   [t_1 F_1^{\dagger} F_d e^{{\it i} (\sqrt{2} - \frac{1}{\sqrt{K}}) \tilde{\phi}_{s,2} } 
e^{{\it i} \sqrt{2}\tilde{\phi}_{s,1} } \\ \nonumber
&+& 
t_2 F_2^{\dagger} F_d e^{ {\it i} (\sqrt{2} - \frac{1}{\sqrt{K}}) \tilde{\phi}_{s,1} }
e^{{\it i} \sqrt{2}\tilde{\phi}_{s,2} } ] S^{-} +H.c. \\ \nonumber
&-& (\sqrt{2} - \frac{1}{\sqrt{K}}) 
( \partial_x \tilde{\phi}_{s,1} +  \partial_x \tilde{\phi}_{s,2} )S_z
\label{tilde-H-t}
\end{eqnarray}
 Note that there are additional phase factors 
 $e^{{i} (\sqrt{2} - \frac{1}{\sqrt{K}}) \tilde{\phi}_{s,\alpha}}$ 
in the hopping terms. 
Since we are interested in the physics close to the 
 localized-delocalized transition, i.e., $K=\alpha^{\ast}\rightarrow1/2$, we may drop these phase factors 
 in the following analysis. The chemical potential term after the above two transformations now becomes
\begin{eqnarray}
\tilde{H}_{\mu}^{''} &=& U_2^{\dagger}U_1^{\dagger} 
\tilde{H}_{\mu} U_1 U_2\\ \nonumber 
&=& -\frac{V}{2} \sqrt{\frac{1}{2K}} 
[ \partial_x (\sqrt{2}\tilde{\phi}_{s,1}) -  
  \partial_x (\sqrt{2}\tilde{\phi}_{s,2})] 
\label{tilde-H-mu}
\end{eqnarray}
Note that since the hoping $\tilde{H}^{"}$ and chemical potential 
$\tilde{H}_\mu^{"}$ terms involve only $\tilde{\phi}_{s,\alpha}$ 
fields, $\tilde{\phi}_{a,\alpha}$ fields are decoupled from 
the Hamiltonian. 
 
Now, we can refermionize the bosons and map our 
transformed Hamiltonian 
\begin{equation}
\tilde{H}_{RLM} \equiv  
\tilde{H}_{bath} + \tilde{H}_t^{''} + \tilde{H}_{\mu}^{''}
\label{maped-RLM}
\end{equation} 
onto the anisotropic Kondo model in Eq.~(\ref{Hkondo}) 
via the following identifications:
\begin{eqnarray}
-\sqrt{2}\tilde{\phi}_{s,1} &=& 
\Phi_{L}^{\uparrow} - \Phi_{R}^{\downarrow}\\ \nonumber
-\sqrt{2}\tilde{\phi}_{s,2} &=& 
\Phi_{R}^{\uparrow} - \Phi_{L}^{\downarrow}\\ \nonumber
{c}_{L/R}^{\sigma}(0) &=& F_{L/R}^{\sigma} e^{{\it i} \Phi_{L/R}^{\sigma} }
\label{refermionization}
\end{eqnarray}
where  $F_{L/R}^{\sigma}$ is the Klein factor for the effective lead $L$ and $R$, respectively.\\

To see the equivalence between these two models, 
we bosonize Eq.~(\ref{Hkondo}) and compare it with Eq.~(\ref{maped-RLM}):

\begin{eqnarray}
H_K &=& H_{leads} + H_{J_\perp} + H_{J_z},\nonumber \\
H_{leads} &=& \sum_{k,\gamma =L,R,\sigma =\uparrow ,\downarrow }[\epsilon
_{k}-\mu _{\gamma }]c_{k\gamma \sigma }^{\dagger }c_{k\gamma \sigma }\nonumber \\
&=& \frac{1}{2} \int dx \sum_{\alpha=L,R}
[(\partial_x \Phi_\alpha)^2(x,t) + \Pi_\alpha^2(x,t)]\nonumber \\
&-&  \frac{V}{2} \sqrt{\frac{1}{2K}} 
\sum_{\sigma=\uparrow,\downarrow} 
[ \partial_x \Phi_{L}^{\sigma} -  
  \partial_x \Phi_{R}^{\sigma}],\nonumber \\ 
H_{J_{\perp}} &=& J_{\perp }^{(1)}s_{LR}^{+}S^{-}+J_{\perp
}^{(2)}s_{RL}^{+}S^{-}+h.c.\nonumber \\
&=&   J_{\perp }^{(1)} F_L^{\dagger \uparrow} F_R^{\downarrow}
e^{{\it i}\Phi_{L}^{\uparrow} - {\it i}\Phi_{R}^{\downarrow}}
+  J_{\perp }^{(2)} F_R^{\dagger \uparrow} F_L^{\downarrow}
e^{{\it i}\Phi_{R}^{\uparrow} - {\it i}\Phi_{L}^{\downarrow}}, \nonumber \\
H_{J_z} &=& \sum_{\gamma =L,R}J_{z}s_{\gamma \gamma
}^{z}S^{z}\nonumber \\
&=& -J_z \sum_{\alpha=L,R} 
[\partial_x \Phi_{\alpha}^{\uparrow} - \partial_x \Phi_{\alpha}^{\downarrow}]\nonumber \\
\label{compare}
\end{eqnarray}
With the proper redefinitions of the Klein factors: 
$F_1^{\dagger} F_d \equiv F_L^{\dagger\uparrow} F_R^{\downarrow}$, 
$F_1^{\dagger} F_d \equiv F_R^{\dagger\uparrow} F_L^{\downarrow}$, 
and the identifications: $d=S^{-}$, $d^{\dagger} = S^{+}$, 
$d^{\dagger}d-1/2 = S_z$, $J_{\perp}^{( \alpha )}= t_{\alpha}$, 
$J_z = 1- \frac{1}{\sqrt{2K}}$,    
we finally establish the equivalence between 
a Kondo model with the effective left ($L$) and 
right lead ($R$) in Eq.~(\ref{Hkondo}) and a dissipative resonant level 
model in Eq.~(\ref{maped-RLM}).

\subsection{Mapping  a dissipative resonant level model 
onto a resonant level coupled to FQHE}

We provide details here on the mapping of 
a dissipative resonant level model Eq.~(\ref{RLM}) 
onto a resonant level coupled to Fractional Quantum Hall Edge states (FQHE) 
as shown in Eq.~(\ref{dot-luttinger}).

We start from the Hamiltonian Eq.~(\ref{dot-luttinger}) describing 
a resonant level coupled to two FQHE states:
\begin{equation}
H_{FQHE} = H_{chiral} + H_{t} + H_{\mu}, 
\label{dot-luttinger-appendix}
\end{equation}
where the lead term $H_{chial}$ describes two chiral Luttinger liquid 
leads with lead index $\alpha=1,2$, $H_t$ denotes the tunneling term and the 
bias voltage term $H_{\mu}$ is given respectively by:  
\begin{eqnarray}
H_{chiral} &=& \frac{1}{2}\int_{-\infty}^{+\infty} 
\sum_{\alpha=1,2} \left(\frac{d\varphi_{\alpha}}{dx}\right)^2  dx,\nonumber \\
H_t &=& t_1 e^{{\it i}\varphi_1/\sqrt{K}} d + t_2 e^{{\it i}\varphi_2/\sqrt{K}} d + h.c.
\nonumber \\
H_{\mu} &=& -\frac{V}{2}\frac{1}{\sqrt{K}} (\partial \varphi_{1} - \partial \varphi_{2}),\nonumber \\
\label{H-LL-appendix}
\end{eqnarray}
where the boson field $\varphi_{\alpha=1,2}$ denotes the chiral Luttinger 
liquid in lead $\alpha$, the tunneling between lead and the resonant level 
is given by $t_{\alpha}$, $V$ is the bias voltage, and 
$K$ refers to the Luttinger parameter. 

Via similar Unitary transfermations mentioned above, 
 ${U}_1 = e^{{\it i} (\frac{\varphi_{1}}{\sqrt{K}} 
- \sqrt{2} \varphi_{1} )S_z }$ and 
${U}_2 = e^{{\it i} (\frac{\varphi_{2}}{\sqrt{K}} 
- \sqrt{2} \varphi_{2} ) S_z }$, Eq.~(\ref{dot-luttinger}) 
now becomes:
\begin{equation}
\bar{H}_{FQHE} = U_2^{\dagger} U_1^{\dagger}H_{FQHE}U_1 U_2 = H_{chiral} + \bar{H}_t + \bar{H}_{\mu},
\label{barH-FQHE}
\end{equation}
where the tunneling 
term $H_t$ in Eq.~(\ref{H-LL-appendix}) becomes (assumming $t_1=t_2=t$):
\begin{eqnarray} 
\bar{H}_t &=&  
t [e^{{\it i} (\sqrt{2} - \frac{1}{\sqrt{K}}) \varphi_{2}} e^{{\it i} \sqrt{2}\varphi_{1}} \\ \nonumber 
&+& e^{{\it i} (\sqrt{2} - \frac{1}{\sqrt{K}}) \varphi_{1}} e^{{\it i} \sqrt{2}\varphi_{2}} ] S^{-} + h.c.\\ \nonumber
&-& (1 - \sqrt{\frac{1}{2 K}}) (\partial 
\sqrt{2}\varphi_{1} + \partial \sqrt{2} \varphi_2) 
S_z, 
\end{eqnarray}
and the chemical potential term in Eq.~(\ref{H-LL-appendix}) becomes
\begin{eqnarray}
\bar{H}_\mu &=& -\frac{V}{2} \sqrt{\frac{1}{2K}} 
[ \partial_x (\sqrt{2}\varphi_{1}) -  
  \partial_x (\sqrt{2}\varphi_{2})]. 
\end{eqnarray}

The equivalence between a resonant level coupled to FQHE 
Eq.~(\ref{dot-luttinger}) and a dissipative resonant level model 
Eq.~(\ref{RLM}) is established by 
comparing the transformed Hamiltonian $\bar{H}_{FQHE}$ in Eq.~(\ref{barH-FQHE})
 for the former model and $\tilde{H}_{RLM}$ 
(see Eq.~(\ref{tilde-H-bath}), (\ref{tilde-H-t}), and (\ref{tilde-H-mu})) 
for the latter one.

\subsection{\bf Mapping a dissipative resonant level onto a 
dissipative resonant level coupled to chiral Luttinger liquid leads }

Below we provide details on the mapping of 
a large dissipative resonant level onto a large 
resonant level (spinless quantum dot) 
with Ohmic dissipation coupled to two chiral 
Luttinger liquid leads. The mapping is easily extended to 
the latter case with a small (single-level) resonant level. 

First, we take the same dissipative boson environment as shown in 
Eq.~(\ref{RLM-diss}). 
Via standard bosonization (see Eq.~(\ref{bosonization-LL})), 
the Luttinger leads and the chemical potential term take the same 
bosonized form as Eq.~(\ref{RLM-leads}) and 
Eq.~(\ref{H-LL}), respectively. 
The remaining parts of the Hamiltonian 
are modified as follows:
\begin{eqnarray}
H_{dot} &=& H_d + H_t + H_{db} ,\nonumber \\
H_d &=& \sum_k \epsilon_{d_k} d^{\dagger}_k d_k, \nonumber \\
H_t &=& \sum_{k,k^{\prime},\alpha=1,2} t_{\alpha}
c^{\dagger}_{k,\alpha} d_{k^{\prime}} S^{-} + h.c., \\
H_{db} &=& \sum_{r,k^{\prime}} \lambda_{r} (d^{\dagger}_{k^{\prime}} d_{k^{\prime}}-1/2) 
(b_{r} + b^{\dagger}_{r}),  \nonumber
\end{eqnarray}
where $\epsilon_{d_k}$ referrs to the energy spectrum of the many-levl 
dot, the electron destruction operator on the dot $d(0)$ is defined as: 
$d(0)= \sum_{k}d_k$, and spin-flip operator $S^{\pm}$ represents for the 
hoping of charge between lead and the dot\cite{lehur1}. We then bosonize the 
electron operators in the leads (see Eq.~(\ref{bosonization-LL}))  
and on the dot: 
$d(0) = \frac{1}{\sqrt{2\pi a}} F_{d}  e^{{\it i}\phi_d}$. Via the unitary 
transformation $U_B$ defined in Eq.~\ref{Ub}, we arrive at:
\begin{eqnarray}
\tilde{H_t} &=& U_B^{\dagger} H_t U_B \\ \nonumber
&=& \sum_{\alpha=1,2} t_{\alpha}  F_{\alpha}^{\dagger} F_d
e^{{\it i} \sqrt{\frac{1}{K_b}} \hat{\phi}_0 } 
e^{{\it i} (\frac{\varphi_{\alpha}(0)}{\sqrt{K}}-\phi_d)} S^{-} + H.c. 
\label{Ubp}
\end{eqnarray}
To further simplify the hoping term, we define new boson fields 
$\phi_{s(a),\alpha}$  
via linear combinations of the fields $(\varphi_{\alpha}(0)$ and $\phi_d)$:
\begin{eqnarray}
\phi_{a,\alpha} &=& \sqrt{K^{\prime}}( 
\frac{\varphi_{\alpha}(0)}{\sqrt{K}}-\phi_d),\nonumber \\
\phi_{s,\alpha} &=& \sqrt{K^{\prime}} 
(\frac{\varphi_{\alpha}(0)}{\sqrt{K}}+\phi_d)
\end{eqnarray}
with $\frac{1}{K^{\prime}}= \frac{1}{K} +1$. 
The combined fermionic baths of the leads and the dot are given by:
\begin{eqnarray}
H_f &\equiv& H_{leads} +H_{d} \nonumber \\
&=& \frac{1}{2} \int dx \sum_{\alpha=1,2}
[(\partial_x \phi_{s,\alpha})^2(x,t) 
+ \tilde{\Pi}_{s,\alpha}^2(x,t)\nonumber \\
&+& (\partial_x \phi_{a,\alpha})^2(x,t) 
+ \tilde{\Pi}_{a,\alpha}^2(x,t)],
\label{Kprime}
\end{eqnarray}
where $\tilde{\Pi}_{s(a),\alpha}$ are canonically conjugate boson fields to 
$\phi_{s(a),\alpha}$ fields. 
In terms of the new fields $\phi_{s(a),\alpha}$, 
the hoping and chemical potential terms now become:
\begin{eqnarray}
\tilde{H}_t &=& \sum_{\alpha=1,2} t_{\alpha}  F_{\alpha}^{\dagger} F_d
e^{{\it i} \sqrt{\frac{1}{K_b}} \hat{\phi}_0 } 
e^{{\it i} \frac{\phi_{a,\alpha}}{\sqrt{K^{\prime}}}} S^{-} + H.c.,\nonumber\\ 
H_{\mu}&\rightarrow& \tilde{H}_{\mu}\nonumber \\ 
&=& -\frac{V}{2} \sqrt{\frac{K}{K^{\prime}}} 
[ \partial_x (\phi_{a,1}) -  
  \partial_x (\phi_{a,2})]. 
\end{eqnarray}
We may furthermore combine the boson 
fields from the leads $\phi_{a,\alpha}$ and from the dissipative bath 
$\hat{\phi}_0$ via the following definitions:
\begin{eqnarray}
\tilde{\phi}_{s,\alpha} &=&\sqrt{\tilde{K}} 
(\frac{\phi_{a,\alpha}}{K^{\prime}} + 
\sqrt{\frac{1}{K_b}}\hat{\phi}_0),\nonumber \\
\tilde{\phi}_{a,\alpha} &=&\sqrt{\tilde{K}} 
(\frac{\phi_{a,\alpha}}{K^{\prime}} - \sqrt{\frac{1}{K_b}}\hat{\phi}_0), 
\label{Ktilde}
\end{eqnarray}
where $\frac{1}{\tilde{K}} = \frac{1}{K^{\prime}} + \frac{1}{K_b}$. 
Upon applying the unitary transformation, the combined fermionic and bosnoic 
baths terms become:
\begin{eqnarray}
H_f &=& \frac{1}{2} \int dx \sum_{\alpha=1,2}
[(\partial_x \tilde{\phi}_{s,\alpha})^2(x,t) 
+ \bar{\Pi}_{s,\alpha}^2(x,t)\nonumber \\
&+& (\partial_x \tilde{\phi}_{a,\alpha})^2(x,t) 
+ \bar{\Pi}_{a,\alpha}^2(x,t)],
\end{eqnarray}
where $\bar{\Pi}_{s(a),\alpha}$ are canonically conjugate boson fields to 
$\tilde{\phi}_{s(a),\alpha}$ fields. 

Meanwhile, the corresponding hoping and chemical potential terms become: 
\begin{eqnarray}
\tilde{H}_t &=& \sum_{\alpha=1,2} t_{\alpha}  F_{\alpha}^{\dagger} F_d
e^{{\it i} \frac{\tilde{\phi}_{s,\alpha}}{\sqrt{\tilde{K}}}} S^{-} + H.c., \\ 
H_{\mu}&\rightarrow& \tilde{H}_{\mu}\nonumber \\ 
&=& -\frac{V}{2} \sqrt{\frac{K^{\prime}}{\tilde{K}}} 
[ \partial_x (\tilde{\phi}_{s,1}) -  
  \partial_x (\tilde{\phi}_{s,2})]. 
\end{eqnarray}
Via the similar Unitary transfermation mentioned above, 
 ${U}_1 = e^{{\it i} (\frac{\tilde{\phi}_{1}}{\sqrt{\tilde{K}}} 
- \sqrt{2} \varphi_{1} )S_z }$ and 
${U}_2 = e^{{\it i} (\frac{\tilde{\phi}_{2}}{\sqrt{\tilde{K}}} 
- \sqrt{2} \tilde{\phi}_{s,2} ) S_z }$, 
the tunneling 
term becomes (assumming $t_1=t_2=t$):
\begin{eqnarray} 
H_t &=&  
t [e^{{\it i} (\sqrt{2} - \frac{1}{\sqrt{\tilde{K}}}) \tilde{\phi}_{s,2}} 
e^{{\it i} \sqrt{2}\tilde{\phi}_{s,1}} \\ \nonumber 
&+& e^{{\it i} (\sqrt{2} - \frac{1}{\sqrt{\tilde{K}}}) \varphi_{1}} 
e^{{\it i} \sqrt{2}\tilde{\phi}_{s,2}} ] S^{-} + h.c.\\ \nonumber
&-& (1 - \sqrt{\frac{1}{2 \tilde{K}}}) (\partial 
\sqrt{2}\tilde{\phi}_{s,1} + \partial \sqrt{2} \tilde{\phi}_{s,2}) 
S_z. 
\end{eqnarray}
The chemical potential term therefore becomes
\begin{eqnarray}
H_{\mu} &\rightarrow& \tilde{H}_{\mu} \nonumber \\
&=& -\frac{V}{2} \sqrt{\frac{K^{\prime}}{2\tilde{K}}} 
[ \partial_x (\sqrt{2}\tilde{\phi}_{s,1}) -  
  \partial_x (\sqrt{2}\tilde{\phi}_{s,2})]. 
\end{eqnarray}
We may now follow the same refermionization procedure as
shown in Eq.~(\ref{refermionization}) to map our Hamiltonian 
onto the anisotropic Kondo model in the same form as 
Eq.~(\ref{Hkondo}) with the following identifications:
\begin{eqnarray}
J_{\perp}^{(1),(2)} &\propto& t_{\alpha} e^{{\it i} 
(\sqrt{2} - \frac{1}{\tilde{K}})\tilde{\phi}_{s;2,1}},\nonumber \\
J_z&\propto& 1-\frac{1}{2\tilde{K}},\nonumber \\
\mu &\rightarrow&\tilde{\mu} = \frac{V}{2} \sqrt{\frac{K^{\prime}}{\tilde{K}}}. 
\end{eqnarray}
The above mapping can easily be generalized to a small quantum dot with 
single resonant level with $\tilde{K}$ given by Eq.~(\ref{K-small}) where 
the contribution from the many-level big dot is absent here.

\section{Average currents.}

In this Appendix, we prove that the average currents in the original 
model $\hat{I}_{ori}$ is equivalent to that in the effective Kondo model 
$\hat{I}_{Kondo}$.
The current operators in both models are given by:
\begin{eqnarray}
\hat{I}_{ori} &=& d/dt (N_1 - N_2) \\ \nonumber
 &=& {\it i} t_1 \sum_k (c^{\dagger}_{k1} d - d^{\dagger} c_{k1}) - 
                                             (1\rightarrow 2) 
\end{eqnarray}
\begin{eqnarray}
\hat{I}_{Kondo} &=& d/dt (N_L - N_R) \\ \nonumber 
 &=& {\it i} J_{\perp}^{(1)} (s_{LR}^{-} S^{+} - s_{RL}^{+} S^{-}) - 
                                        (1\rightarrow 2, L\rightarrow R)
\label{Ikondo}
\end{eqnarray}

On the other hand, from the bosonized forms of the two models, (at the transition) we have:

\begin{eqnarray}
\langle \hat{I}_{ori}\rangle &=& \langle d/dt (N_1 - N_2)\rangle \\ \nonumber 
                                  &=& \int\ dx \langle 
\frac{d}{dt} [ \partial_x \varphi_{1} -  \partial_x \varphi_{2}] \rangle  \\ \nonumber
                                  &=& \int \ dx\sqrt{\frac{1}{2K}} \langle 
\frac{d}{dt} [ \partial_x (\sqrt{2}\tilde{\phi}_{s,1}) -  
                                                      \partial_x (\sqrt{2}\tilde{\phi}_{s,2})]  \rangle
\end{eqnarray}

\begin{eqnarray}
\langle \hat{I}_{Kondo}\rangle &=& \langle  d/dt (N_L - N_R)\rangle  \\ \nonumber
 &=& \int dx\  \langle d/dt \sum_{\sigma=\uparrow,\downarrow} 
 [ \partial_x \Phi_{L}^{\sigma} -  \partial_x \Phi_{R}^{\sigma}] \rangle  \\ \nonumber
 &=& \int dx\ \langle d/dt[ \partial_x (\sqrt{2}\tilde{\phi}_{s,1}) -  
                                                             \partial_x (\sqrt{2}\tilde{\phi}_{s,2})] \rangle
\end{eqnarray}

Therefore, we have 
\begin{equation}
\langle \hat{I}_{ori}\rangle \ = \frac{1}{\sqrt{2K}} \langle \hat{I}_{Kondo}\rangle
\end{equation}
(or $\langle\hat{I}_{ori}\rangle \ = \frac{1}{\sqrt{2\alpha^{\ast}}} \langle\hat{I}_{Kondo}\rangle$). 
The above relation obtained so far from the mapping is exact at finite bias voltages. 
In the limit of our interest $K=\alpha^{\ast} \rightarrow 1/2$, 
$\langle \hat{I}_{ori}\rangle \ =\ \langle\hat{I}_{Kondo}\rangle$.\\

We can also prove this equivalence through Keldysh perturbation theory.
We now would like to prove that 
\begin{equation}
\langle\hat{I}_{Kondo}\rangle_K = \langle\hat{I}_{ori}\rangle_{ori}
\end{equation}
where 
\begin{eqnarray}
\langle I_{Kondo}(t)\rangle_K &=& 
\frac{1}{Z_K} \\ \nonumber
&\times& Tr[e^{-\beta H_{K}} T_c(S_c^{K}(-\infty,\infty) I_{Kondo}(t))]
\nonumber \\
Z_{K} &=& Tr[e^{-\beta H_{K}} T_c(S_c^{K}(-\infty,\infty))]\nonumber \\
S_c^{K}(-\infty,\infty) &=& e^{-{\it i} \int_c dt' H_{K}^{eq}(t')}
\end{eqnarray}
and 
\begin{eqnarray}
\langle\hat{I}_{ori}(t)\rangle_{ori} &=& 
\frac{1}{Z_{ori}} Tr[e^{-\beta H} T_c(S_c^{ori}(-\infty,\infty) \hat{I}_{ori}(t))]
\nonumber \\
Z_{ori} &=& Tr[e^{-\beta H} T_c(S_c^{ori}(-\infty,\infty))]\nonumber \\
S_c^{ori}(-\infty,\infty) &=& e^{-{\it i} \int_c dt' H^{eq}(t')}
\end{eqnarray}

Here $H_{K}^{eq}(H^{eq}) $ is the Kondo (original) 
Hamiltonian in equilibrium ($\mu =0$), 
$T_c(\cdots)$ orders the operators along the Keldysh contour $c$.

1. We first show that $Z_{K} = Z_{ori}$ (the two partition functions 
from the original and the effective Kondo models are equivalent) where 
\begin{eqnarray}
Z_{ori} &=& Tr[e^{-\beta H} T_c(S_c^{ori}(-\infty,\infty))]\\ \nonumber
S_c^{ori}(-\infty,\infty) &=& e^{-{\it i} \int_c dt' H^{eq}(t')}
\end{eqnarray}

To prove this, we first note that the original and 
the effective Kondo models are related by the above-mentioned 
unitary transfromations: $H_K = U^{\dagger} H U$ with $U = U_2 U_1 U_B$. 
The similar relation holds for the current operators: 
$\hat{I}_{Kondo} =  U^{\dagger} \hat{I}_{ori} U$. 
Using the following identities:
\begin{eqnarray}
\label{identity}
\noindent
&e&^{-\int_0^{\beta} d\tau U^{\dagger}(\tau) H(\tau) U(\tau)} \\ \nonumber
&=&  \sum_{n=0}^{\infty} \frac{(-1)^n}{n!} [\int_0^{\beta} d\tau 
(U^{\dagger}(\tau) H(\tau) U(\tau)]^n
\end{eqnarray}
where 
\begin{eqnarray}
Tr[\hat{A}(\tau) \hat{B}(\tau) \hat{C}(\tau)] &=& 
Tr[\hat{C}(\tau) \hat{A}(\tau) \hat{B}(\tau)]\nonumber \\
 &=& Tr[\hat{B}(\tau) \hat{C}(\tau) \hat{A}(\tau)] =\cdots
\end{eqnarray}
with $\tau={\it i} t$ the imaginary time and 
$\hat{A}, \hat{B}, \hat{C}$ being any quantum mechanical operators  
we can then show that $Z_{K} = Z_{ori}$. In other words, 
when the Hamiltonian is under above unitary transformations, 
the trace in the partition function remains unchanged.

2. In the similar way, we can prove that:
\begin{eqnarray}
 & & Tr[e^{-\beta H_{K}} T_c(S_c^{K}(-\infty,\infty) I_{Kondo}(t))] \\ \nonumber
 &= & Tr[e^{-\beta H} T_c(S_c^{ori}(-\infty,\infty) I_{ori}(t))]
\end{eqnarray}
where we have used Eq.~(\ref{identity}) and $\hat{I}_{Kondo} =  U^{\dagger} \hat{I}_{ori} U$. 

From 1. and 2. mentioned above, we conclude that 
$\langle\hat{I}_{Kondo}(t)\rangle_K = \langle\hat{I}_{ori}(t)\rangle_{ori}$ 
holds for all orders in Keldysh perturbation theoy.\\

\section{Non-equilibrium current for $t_1 \neq t_2$.}

In this Appendix, we derive the general expression for the average 
current for $t_1 \neq t_2$. From Eq.~(\ref{Ikondo}), 
the average current in the Kondo model is given by:
\begin{equation}
\langle\hat{I}\rangle = \int_{-\infty}^{\infty} \sum_k J_{\perp}^{(1)} 
({\it G}_{k,d}^{<}(\omega) -{\it G}_{d,k}^{<}(\omega)) -(1\rightarrow 2, L\rightarrow R) 
\end{equation}
where ${\it G}_{k,d}^{<}(t) = {\it i} \langle s_{LR}^{-} S_d(t)\rangle$. 
Following Ref.\onlinecite{Meir}, the Dyson's equation for  
${\it G}_{k,d}^{<}(\omega)$ is given by:
\begin{eqnarray}
{\it G}_{k,d}^{<}(\omega) &=& J_{\perp}^{(1)}  [{\chi_{LR}^{+-}}^t(\omega)  
{\chi_d^{+-}}^<(\omega)  \\ \nonumber 
&-&
{\chi_{LR}^{+-}}^<(\omega)  
{\chi_d^{+-}}^{\tilde{t}}(\omega)] -(1\rightarrow 2, L\rightarrow R)
\end{eqnarray}
where ${\chi_{LR}^{+-}}^<(t) = \langle s_{LR}^- s_{LR}^+(t)\rangle$, 
${\chi_{LR}^{+-}}^<(t) = \langle s_{LR}^-(t) s_{LR}^+\rangle$, 
 ${\chi_d^{+-}}^< = \langle S_d^- S_d^+(t)\rangle$, 
${\chi_{LR}^{+-}}^t(\omega)$, and  
 ${\chi_d^{+-}}^{\tilde{t}}$ are 
time-order and anti-timeordered Green's functions, respectively. 
The following relations hold among these correlation functions: 
\begin{eqnarray}
\chi^<(\omega) +\chi^>(\omega) &=& \chi^t(\omega) + 
\chi^{\tilde{t}}(\omega)\\ \nonumber
\chi^>(\omega) -\chi^<(\omega) &=& \chi^R(\omega) - 
\chi^{A}(\omega)\\ \nonumber
\end{eqnarray}
where $\chi^{R/A}(\omega)$ is the retarded (advanced) 
Green's function, respectively. Straightforward calculation gives: 
\begin{eqnarray}
{\chi_{LR}^{+-}}^<(\omega) &=& 2\pi f_{\omega-\mu_L} (1-f_{\omega-\mu_R}) 
\delta(\omega-\epsilon(k))\\ \nonumber 
{\chi_{LR}^{+-}}^>(\omega) &=& -2\pi f_{\omega-\mu_R} (1-f_{\omega-\mu_L}) 
\delta(\omega-\epsilon(k))
\end{eqnarray}
The average current reads
\begin{eqnarray}
\langle\hat{I}\rangle &=& \int_{-\infty}^{\infty} [f_{\omega-\mu_L} (1-f_{\omega-\mu_R}) 
\tilde{\Gamma}_1 \\ \nonumber 
&-& f_{\omega-\mu_R} (1-f_{\omega-\mu_L}) 
\tilde{\Gamma}_2] (\chi_d^R(\omega) - \chi_d^A(\omega))\\ \nonumber
&+& 
(\tilde{\Gamma}_1-\tilde{\Gamma}_2) \chi_d^<(\omega)
\end{eqnarray}
where $\tilde{\Gamma}_{1,2} = 2\pi \rho_0 (J_{\perp}^{(1),(2)})^2$ with $\rho_0$ being 
the constant density of states of the leads. 

Following Ref.~\onlinecite{Meir}, 
for $\tilde{\Gamma}_1 = \lambda \tilde{\Gamma}_2$, we have 
\begin{equation}
\langle\hat{I}\rangle\ = \int_{-\infty}^{\infty} (f_{\omega-\mu_L} -f_{\omega-\mu_R}) 
\tilde{\Gamma}(\omega) (\chi_d^R(\omega) - \chi_d^A(\omega)) 
\end{equation}
where $\tilde{\Gamma}(\omega) = (2\pi \rho_0)^2 
\frac{(g_{\perp}^1 (\omega)g_{\perp}^2 (\omega))^2}
{(g_{\perp}^1 (\omega))^2 + (g_{\perp}^2 (\omega))^2}$. 
Note that the Kondo couplings have been genralized to be 
frequency dependent following the noneuilibrium RG approach.


\begin{references}

\bibitem{sachdevQPT}
S. Sachdev, {\it Quantum Phase Transitions}, Cambridge University Press (2000).

\bibitem{Steve}
S. L. Sondhi, S. M. Girvin, J. P. Carini, and D. Shahar, 
Rev. Mod. Phys. {\bf 69}, 315 (1987).

\bibitem{vojtaRMP}
H. v. L\"ohneysen, A. Rosch, M. Vojta, and P. W\"olfle, 
Rev. Mod. Phys. {\bf 79}, 1015 (2007). 

\bibitem{saleur}
I. Safi and H. Saleur, Phys. Rev. Lett. {\bf 93}, 126602 (2004). 

\bibitem{Ingold}
M.H. Devoret, D. Esteve, H. Grabert, G.-L. Ingold, H. Pothier, and C. Urbina, 
Phys. Rev. Lett. {\bf 64}, 1824 (1990).

\bibitem{nazarov-book}
G. Ingold and Yu. V. Nazarov, in Single Charge Tunneling, 
edited by H. Grabert and M. H. Devoret (Plenum Press, New York, 1992). 

\bibitem{lehur1}
K. Le Hur, Phys. Rev. Lett. {\bf 92}, 196804 (2004);
M.-R. Li, K. Le Hur, and W. Hofstetter, Phys. Rev. Lett. {\bf 95}, 086406 (2005).

\bibitem{lehur2}
K. Le Hur and M.-R. Li, Phys. Rev. B {\bf 72}, 073305 (2005).

\bibitem{zarand}
 L. Borda, G. Zarand, P. Simon, Phys. Rev. B {\bf 72}, 155311 (2005); 
L. Borda, G. Zarand, and D. Goldhaber-Gordon, cond-mat/0602019 (un-published).


\bibitem{Markus}
P. Cedraschi and M. B\" uttiker, Annals of Physics (NY) {\bf 289}, 1 (2001).

\bibitem{matveev}
A. Furusaki and K. A. Matveev, Phys. Rev. Lett. {\bf88}, 226404 (2002).

\bibitem{gefen} 
M. Goldstein, Y. Gefen, and R. Berkovits, 
Phys. Rev. B {\bf 83}, 245112 (2011). 

\bibitem{buttiker}
P. Cedraschi, V. V. Ponomarenko, and M. B\"uttiker, Phys. Rev. Lett. 
{\bf 84}, 346 (2000).
        
\bibitem{Zarand2}
G. Zarand, C.H. Chung, P. Simon, and M. Vojta, 
Phys. Rev. Lett. {\bf 97}, 166802 (2006).

\bibitem{zoller}
A. Recati, P. O. Fedichev, W. Zwerger, J. von Delft, and P. Zoller, 
Phys. Rev. Lett. {\bf 94}, 040404 (2005).


\bibitem{chung1}
C.H. Chung, K. Le Hur, M. Vojta and P. W\"olfle, Phys. Rev. Lett {\bf 102}, 216803 (2009).

\bibitem{chung2}
C.H. Chung, K.V.P. Latha, K. Le Hur, M. Vojta and P. W\"olfle, 
Phys. Rev. B, {\bf 82}, 115325 (2010).

\bibitem{latha}
C.H. Chung and K.V.P. Latha, Phys. Rev. B {\bf 82}, 085120 (2010).

\bibitem{Feldman}
D. E. Feldman, Phys. Rev. Lett., {\bf 95}, 177201 (2005). 

\bibitem{mitra}
A. Mitra, S. Takei, Y.B. Kim, and A. J. Millis, 
Phys. Rev. Lett., {\bf 97}, 236808 (2006). 

\bibitem{Goldhaber}
R. M. Potok, I. G. Rau, H. Shtrikman, Y. Oreg, and D. Goldhaber-Gordon,
Nature {\bf 447}, 167 (2007).

\bibitem{QMSi}
S. Kirchner, Q.M. Si, Phys. Rev. Lett. {\bf 103}, 206401 (2009).

\bibitem{ybkim}
S. Takei, Y.B. Kim, Phys. Rev. B {\bf 76} 115304 (2007); 
S. Takei, W. Witczak-Krempa, Y.B. Kim, Phys. Rev. B {\bf 81}, 125430 (2010).

\bibitem{zwerger}
Dittrich, P. Hanggi, G-I. Ingold, B. Kramer and G. Sch\"on, W. Zwerger, 
Quantum Transport and Dissipation (Viley-VCH, 1998).

\bibitem{florens}
S. Florens, P. Simon, S. Andergassen1, and D. Feinberg, Phys. Rev. B {\bf 75}, 155321 (2007).

\bibitem{keldysh}
J. Schwinger, J. Math. Phys. {\bf 2}, 407 (1961); 
L. V. Keldysh, Soviet Physics JETP {\bf 20}, 1018 (1965); 
J. Rammer and H. Smith, Rev. Mod. Phys. {\bf 58}, 323 (1986); 
A. P. Jauho, N. S. Wingreen and Y. Meir, Phys. Rev. B {\bf 50}, 5528 (1994).

\bibitem{noneqRG} 
A. Rosch J. Paaske, J. Kroha and P. W\"olfle, 
Phys. Rev. Lett. {\bf 90}, 076804 (2003); 
A. Rosch, J. Paaske, J. Kroha, P. W\"olfle, 
J. Phys. Soc. Jpn. {\bf 74}, 118 (2005).

\bibitem{schoellerRG}
H. Schoeller, Eur. Phys. J. Special Topics {\bf 168}, 179 (2009).

\bibitem{kehrein}
 S. Kehrein, Phys. Rev. Lett. {\bf 95}, 056602 (2005).

\bibitem{meden}
S. G. Jakobs, V. Meden and H. Schoeller, Phys. Rev. Lett. {\bf 99}, 
150603 (2007); C. Karrasch et al., Phys. Rev. B {\bf 81}, 125122 (2010).

\bibitem{FRG}
H. Schmidt and P. W\"olfle, Ann. Phys. (Berlin) {\bf 19}, 60 (2010).

\bibitem{mora}
C. Mora, P. Vitushinsky, X. Leyronas, A. A. Clerk, and K. Le Hur, 
Phys. Rev. B {\bf 80}, 155322 (2009); 
P. Vitushinsky, A. A. Clerk, and K. Le Hur, Phys. Rev. Lett. {\bf 100}, 
036603 (2008); C. Mora, X. Leyronas and N. Regnault, Phys. Rev. Lett. 
{\bf 100}, 036604 (2008); Z. Ratiani and A. Mitra, Phys. Rev. B {\bf 79}, 
24511 (2009).

\bibitem{timm}
C. Timm, Phys. Rev. B {\bf 83}, 115416 (2011).

\bibitem{fendley}
P. Fendley, A. W. W. Ludwig and H. Saleur, Phys. Rev. Lett. {\bf 74}, 3005 (1995).

\bibitem{andrie}
P. Mehta and N. Andrei, Phys. Rev. Lett. {\bf 96}, 216802 (2006); 
P. Mehta and N. Andrei, arXiv:0702612 (2007).

\bibitem{hur-eq-noneq}
P. Dutt, J. Koch, J.E. Han, K. Le Hur, Annals of Physics {\bf 326}, 2963 (2011).

\bibitem{saleur-dagotto}
E. Boulat, H. Saleur and P. Schmitteckert, Phys. Rev. Lett. {\bf 101}, 
140601 (2008); 
F. Heidrich-Meisner, A. E. Feiguin and E. Dagotto, Phys. Rev. B {\bf 79}, 
235336 (2009).

\bibitem{anders}
F. B. Anders, Phys. Rev. Lett. {\bf 101}, 066804 (2008); 
S. Schmitt and F.B. Anders, Phys. Rev. B {\bf 81}, 165106 (2010).

\bibitem{fabrizio}
M. Schiro and M. Fabrizio, Phys. Rev. B {\bf 79}, 155302 (2009); 
P. Werner, T. Oka and A. J. Millis, Phys. Rev. B {\bf 79}, 035320 (2009); 
T. L. Schmidt, P. Werner, L. Muehibacher, A. Komnik, Phys. Rev. B {\bf 78}, 235110 (2008); 
P. Werner et al., Phys. Rev. B {\bf 81}, 035108 (2010); 
L. Muehlbacher, D. F. Urban, and A. Komnik, 
Phys. Rev. B {\bf 83}, 075107 (2011); 
E. Gull, A. J. Millis, A.I. Lichtenstein, A. N. Rubtsov, M. Troyer, 
and P. Werner, Rev. Mod. Phys. {\bf 83}, 349 (2011) .

\bibitem{han}
J. E. Han and R. J. Heary, Phys. Rev. Lett. {\bf 99}, 236808 (2007); 
J. E. Han, Phys. Rev. B {\bf 81}, 245107 (2010).

\bibitem{Mckenzie2}
J. Gilmore and R. H. McKenzie, 2005 J. Phys.: Condens. Matter {\bf 17}, 
1735.

\bibitem{FQHE}
See, for example: A. Feiguin, P. Fendley, M. Fisher, and C. Nayak, 
Phys. Rev. Lett. {\bf 101}, 236801 (2008); A. M. Chang, Rev. Mod. Phys. {\bf 75}, 1449 (2003); M. Grayson, D. C. Tsui, L. N. Pfeiffer, K. W. West, 
and A. M. Chang, Phys. Rev. Lett. {\bf 86}, 2645 (2001).  

\bibitem{finkelstein}
Y. Bomze, H. Mebrahtu, I. Borzenets, A. Makarovski, and G. Finkelstein, 
Phys. Rev. B {\bf 79}, 241402(R) (2009); H. T. Mebrahtu, 
I. V. Borzenets, D. E. Liu, H. Zheng, 
Y. V. Bomze, A. I. Smirnov, H. U. Baranger, and G. Finkelstein, 
Nature {\bf 488}, 61, (2012).


\bibitem{Josephson}
G. Refael, E. Demler, Y. Oreg, and D. S. Fisher, 
Phys. Rev. B {\bf 75}, 014522 (2007).

\bibitem{demler}
G. D. Torre, E. Demler, T. Giamarchi, E. Altman, 
Nat. Phys. 6, 806-810 (2010); E. G. Dalla Torre, E. Demler, T. Giamarchi, E. Altman, Phys. Rev. B {\bf 85}, 184302 (2012).

\bibitem{lobos}
A. M. Lobos and T. Giamarchi, Phys. Rev. B {\bf 84}, 024523 (2011).

\bibitem{Kapitulnik}
N. Mason and A. Kapitulnik, Phy. Rev. B {\bf 65}, 220505 (2002).

\bibitem{zimanyi}
K.-H. Wagenblast, A. van Otterlo, G. Sch\"on, and G. T. 
Zim\'anyi, Phys. Rev. Lett. {\bf 78}, 1779 (1997). 

\bibitem{Rimberg}
A. J. Rimberg, T. R. Ho, C. Kurdak, J. Clarke, K. L. Campman, and A. C. Gossard
Phys. Rev. Lett. {\bf 78}, 2632 (1997).

\bibitem{Kontos}
M.R. Delbecq, V. Schmitt, F.D. Parmentier, N. Roch, J.J.
Viennot, G. F\'eve, B. Huard, C. Mora, A. Cottet and T. Kontos, 
Phys. Rev. Lett. {\bf 107}, 256804 (2011); K. D. Petersson {\it et al.}, Nature {\bf 490}, 380 (2012); H. Toida {\it et al.}, arXiv:1206.074; S. J. Chorley {\it et al.}, Phys. Rev. Lett. {\bf 108}, 036802 (2012); T. Frey {\it et al.}, 
Phys. Rev. B {\bf 86}, 115303 (2012).

\bibitem{cavities}
O. Astafiev, A.M. Zagoskin, A.A. Abdumalikov Jr., Yu. A. Pashkin, T. Yamamoto, K. Inomata, Y. Nakamura, and J.S. Tsai, Science {\bf 327}, 840 (2010); 
K. Le Hur, Phys. Rev. B {\bf 85}, 140506(R) (2012); I.-C. Hoi, 
C.M. Wilson, G. Johansson, T. Palomaki, B. Peropadre, P. Delsing, 
Phys. Rev. Lett. {\bf 107}, 073601 (2011); M. Goldstein, 
M. H. Devoret, M. Houzet, L. I. Glazman, arXiv:1208.0319; 
M. Marthaler, Y. Utsumi, D. S. Golubev, A. Shnirman, and 
Gerd Sch\"on, Phys. Rev. Lett. {\bf 107}, 093901 (2011); 
P.-Q. Jin, M. Marthaler, J. H. Cole, A. Shnirman, and Gerd Sch\"on, 
Phys. Rev. B {\bf 84}, 035322 (2011); H. Zheng, D. J. Gauthier, and 
H. U. Baranger, Phys. Rev. A {\bf 82}, 063816 (2010); 
H. Zheng, D. J. Gauthier, and H. U. Baranger, 
Phys. Rev. A {\bf 85}, 043832 (2012); 
P. Longo, P. Schmitteckert and K. Busch
Phys. Rev. Lett. {\bf 104}, 023602 (2010); 
J.-T. Shen and S. Fan, Phys. Rev. Lett. {\bf 98}, 153003 (2007); 
J.-T. Shen and S. Fan, Phys. Rev. A {\bf 76}, 062709 (2007); 
A. LeClair, F. Lesage, S. Lukyanov, and H. Saleur, Phys. Lett. A 
{\bf 235}, 203 (1997); R. Konik and A. LeClair, 
Phys. Rev. B {\bf 58}, 1872 (1998).  



\bibitem{McKenzie}
J. Gilmore and R. McKenzie, J. Phys. C. {\bf 11}, 2965 (1999).

\bibitem{giamarchi}
T. Giamarchi, Quantum Physics in One Dimension (Oxford University 
Press, Oxford, 2004).

\bibitem{gogolin}
A. O. Gogolin, A. A. Nersesyan, and A. M. Tsvelik, 
Bosonization and Strongly Correlated Systems (Cam-
bridge University Press, Cambridge, 1998). 

\bibitem{delft}
Jan von Delft and Herbert Schoeller, Ann. Phys. (Berlin) {\bf 7}, 225 (1998).



\bibitem{mapping} 
In the delocalized phase, it would be more judicious to redefine $\Psi \to (t_1 c_1+ t_2 c_2)/\sqrt{t_1^2+t_2^2}$, and then bosonize and refermionize to get the fermions with spins \cite{lehur2}, $\Psi_{\sigma}$. For $V\ll T_K$: $\Psi^{\dagger}_{\uparrow} \Psi_{\downarrow} \leftarrow 1/\sqrt{t_1^2+t_2^2} 
(t_1 c^{\dagger}_{L\uparrow} c_{R\downarrow} + 
t_2 c^{\dagger}_{R\uparrow} c_{L\downarrow})$ and 
$\Psi^{\dagger}_{\sigma} \Psi_{\sigma} \leftarrow (1/\sqrt{t_1^2+t_2^2})\sum_{\alpha=L,R}
t_{\alpha}c^{\dagger}_{\alpha \sigma} c_{\alpha \sigma}$.



\bibitem{decoherence}
J. Paaske, A. Rosch, J. Kroha, and P. W\"olfle, Phys. Rev. B {\bf 70}, 155301 
(2004); J. Paaske, A. Rosch, P. W\"olfle, Phys. Rev. B {\bf 69}, 155330 (2004).

\bibitem{Meir}
N. S. Wingreen and Y. Meir, Phys. Rev. B {\bf 49}, 11040 (1994).


\bibitem{glattli} L. Saminadayar {\it et al.}, Phys. Rev. Lett. {\bf 79}, 2526 (1997); R. de-Piciotto {\it et al.}, Nature (London) {\bf 389}, 162 (1997).

\bibitem{statistics} I. Safi, P. Devillars, and T. Martin, Phys. Rev. Lett. {\bf 86}, 4628 (2001); S. Vishveshwara, {\it ibid} {\bf 91}, 196803 (2003); C. Bena and C. Nayak, Phys. Rev. B {\bf 73}, 155335 (2006).

\bibitem{schiller} A. Schiller and S. Hershfield, Phys. Rev. B {\bf 58}, 14978 (1998).

\bibitem{schoeller} T. Korb, F. Reininghaus, H. Schoeller, and J. K\a"onig, 
Phys. Rev. B {\bf 76}, 165316 (2007).

\bibitem{realtimeRG}
P. Moca, G. Zarand, C.H. Chung, P. Simon, Phys. Rev. B {\bf 83}, 
201303(R) (2011).


\bibitem{safi}
I. Safi, arXiv:0908.4382; I. Safi, C. Bena, and A. Cr\a'epieux, Phys. Rev. B {\bf 78}, 205422 (2008).

\bibitem{footnote}
The FRG approach for the Kondo couplings 
$g_{\perp,z}(\omega)$ used here\cite{chung2} 
(Eq.~(\ref{gpergz}) and Eq.~(\ref{gamma})) 
is somewhat different from that in Ref.~\onlinecite{FRG}: 
In Ref.~\onlinecite{FRG} 
the RG scaling equation for pseudofermion 
self-energy is formulated and is solved self-consistently 
along with the RG scaling equations for the Kondo couplings.  
Here, following Ref.~\onlinecite{chung2} and Ref.~\onlinecite{noneqRG} 
the pseudofermion self energy is 
included self-consistently 
through the decoherence rate $\Gamma(\omega)$ where it 
is computed via the Fermi-Golden rule within the 
renormalized perturbation theory. Nevertheless, we have checked that the 
results on $g_{\perp,z}(\omega)$ via both approaches agree very well.

\bibitem{Stefan}
S. Kirchner and Q. Si, Physica Status Solidi (b), {\bf 247}, 631 (2010).

\bibitem{hur-lifetime-lutt}
K. Le Hur, Phys. Rev. B {\bf 74}, 165104 (2006); K. Le Hur, Phys. Rev. Lett. 
{\bf 95}, 076801 (2005).

\end{references}
\end{document}